\documentclass[a4paper,twocolumn]{quantumarticle}

\pdfoutput=1

\usepackage[utf8]{inputenc}
\usepackage[english]{babel}
\usepackage[T1]{fontenc}
\usepackage{amsmath, amsthm, amssymb, amsfonts, mathtools}    
    \mathtoolsset{showonlyrefs=true}
\usepackage[left]{eurosym}
\usepackage{enumitem}
\usepackage{hyperref}
\addto\extrasenglish{
    
}

\usepackage{tikz}
\usepackage{float}
\usepackage{booktabs}
\usepackage{parskip}
\usepackage[ruled, noend]{algorithm2e}
\usepackage[backend=biber,style=numeric,sorting=none,doi=true,isbn=false,url=false,citestyle=numeric-comp]{biblatex}
\usepackage[acronym]{glossaries}
\usepackage{xstring}
\usepackage{xspace}

\makeatletter 
    \newcommand{\myhypertarget}[1]{\Hy@raisedlink{\hypertarget{#1}{}}}
\makeatother

\newcommand\createacronym[3]{%
    \newacronym{#1}{#2}{#3}%
    \StrDel{#2}{-}[\myString]%
    \StrDel{\myString}{(}[\myString]%
    \StrDel{\myString}{)}[\myString]%
    \StrSubstitute{\myString}{\&}{}[\myString]
    \StrSubstitute{\myString}{+}{p}[\myString]%
    \StrSubstitute{\myString}{/}{-}[\myString]%
    \expandafter\newcommand\csname \myString\endcsname{%
        \myhypertarget{#1}{}%
        \hyperlink{#1}{\gls*{#1}}%
        \xspace%
    }%
    \expandafter\newcommand\csname \myString s\endcsname{%
        \myhypertarget{#1}{}%
        \hyperlink{#1}{\glspl*{#1}}%
        \xspace%
    }%
}

\usepackage{todonotes}
\usepackage{xcolor}
\definecolor{sergio}{RGB}{0, 100, 70}
\definecolor{sergioB}{RGB}{180,240,220}

\definecolor{martin}{rgb}{0,.4,1}

\definecolor{elisabeth}{RGB}{0, 131, 143}

\newcommand{\Binary}{\mathbb{B}}
\newcommand{\Natural}{\mathbb{N}}
\newcommand{\Integer}{\mathbb{Z}}
\newcommand{\Real}{\mathbb{R}}
\newcommand{\vectorContIneq}{h}
\newcommand{\vectorContEq}{b}
\newcommand{\vectorInt}{c}
\newcommand{\vectorCont}{c'}
\newcommand{\lowerBound}{\underline{z}}
\newcommand{\upperBound}{\overline{z}}

\newcommand{\matrixIntEq}{A}
\newcommand{\matrixIntIneq}{G}
\newcommand{\matrixContEq}{A'}
\newcommand{\matrixContIneq}{G'}
\newcommand{\varAux}{\alpha}
\newcommand{\upperVarAux}{\overline{\varAux}}

\newcommand{\vectorDemand}{d}
\newcommand{\vectorShedding}{r}  
\newcommand{\vectorGeneration}{g}   
\newcommand{\vectorGenerationUpper}{\overline{g}}   
\newcommand{\vectorFlow}{f}   
\newcommand{\vectorFlowUpper}{\overline{f}}
\createacronym{milp}{MILP}{mixed-integer linear programming}
\createacronym{mblp}{MBLP}{mixed-binary linear programming}
\createacronym{bnb}{B\&B}{Branch-and-Bound}
\createacronym{bd}{BD}{Benders' Decomposition}
\createacronym{qb}{QBD}{Quantum Benders' Decomposition}
\createacronym{lp}{LP}{linear programming}
\createacronym{ip}{IP}{integer programming}
\createacronym{ilp}{ILP}{integer linear programming}
\createacronym{op}{OP}{original problem}
\createacronym{mp}{MP}{master problem}
\createacronym{sp}{SP}{subproblem}
\createacronym{tnep}{TNEP}{transmission network expansion planning}
\createacronym{res}{RES}{renewable energy sources}
\createacronym{qubo}{QUBO}{quadratic unconstrained binary optimization}
\createacronym{qc}{QC}{quantum computing}
\createacronym{sa}{SA}{simulated annealing}
\createacronym{qa}{QA}{quantum annealing}
\createacronym{qaoa}{QAOA}{quantum approximate optimization algorithm}
\createacronym{vqe}{VQE}{variational quantum eigensolver}
\createacronym{nisq}{NISQ}{noisy intermediate-scale quantum}
\createacronym{grb}{GRB}{Gurobi}
\createacronym{bdqa}{BD-QA}{BD with QA}
\createacronym{sdk}{SDK}{software development kit}
\createacronym{qpu}{QPU}{quantum processing unit}
\createacronym{mm}{MM}{Minorminer}
\createacronym{fx}{FX}{Fixed}
\createacronym{tt}{TT}{Tightest}

\begin{document}

\title{Performance enhancing of hybrid quantum-classical Benders' algorithm for MILP optimization}

\author{Sergio López-Baños}
\email{sergio.lopezbanos@dlr.de}
\affiliation{German Aerospace Center (DLR), Institute of Networked Energy Systems, Carl-von-Ossietzky-Str. 15, 26129 Oldenburg, Germany}
\affiliation{Hamburg University of Technology, Institute for Quantum Inspired and Quantum Optimization, Blohmstraße 15, 21079 Hamburg, Germany} 
\orcid{0009-0001-4873-7437}

\author{Elisabeth Lobe}
\affiliation{German Aerospace Center (DLR), Institute of Software Technology, Lilienthalplatz 7, 38108 Brunswick, Germany}
\orcid{0000-0002-3473-8906}

\author{Ontje Lünsdorf}
\affiliation{German Aerospace Center (DLR), Institute of Networked Energy Systems, Carl-von-Ossietzky-Str. 15, 26129 Oldenburg, Germany}
\orcid{0000-0003-4464-5735}

\author{Oriol Raventós}
\affiliation{German Aerospace Center (DLR), Institute of Networked Energy Systems, Carl-von-Ossietzky-Str. 15, 26129 Oldenburg, Germany}
\orcid{0000-0002-0512-4331}

\maketitle

\begin{abstract}
    Mixed-integer linear programming problems are extensively used in industry for a wide range of optimization tasks. However, as they get larger, they present computational challenges for classical solvers within practical time limits. Quantum annealers can, in principle, accelerate the solution of problems formulated as quadratic unconstrained binary optimization instances, but their limited scale currently prevents achieving practical speedups. Quantum-classical algorithms have been proposed to take advantage of both paradigms and to allow current quantum computers to be used in larger problems. In this work, a hardware-agnostic Benders' decomposition algorithm and a series of enhancements with the goal of taking the most advantage of quantum computing are presented. The decomposition consists of a master problem with integer variables, which is reformulated as a quadratic unconstrained binary optimization problem and solved with a quantum annealer, and a linear subproblem solved by a classical computer. The enhancements consist, among others, of different embedding processes that substantially reduce the pre-processing time of the embedding computation without compromising solution quality, a conservative handling of cut constraints, and a stopping criterion that accounts for the limited size of current quantum computers and their heuristic nature. The proposed algorithm is benchmarked against classical approaches using a D-Wave quantum annealer for a scalable family of transmission network expansion planning problems.

\end{abstract}

\section{Introduction} \label{sec: Introduction}
Quantum computers have gained traction over the last decade in solving combinatorial optimization problems, because the exploitation of quantum mechanical properties might offer an advantage over previously available only classical methods. However, due to the limitations of current quantum hardware and the nature of problems in many domains, quantum computers should be used alongside classical computers in specific subroutines to seek for a computational speedup.

\subsection{Background}
In \QC, each binary variable is encoded in a two-state quantum system, whose quantum properties enable computational strategies that are fundamentally different from classical approaches. However, the limitation of current \NISQ hardware~\cite{Preskill2018}, particularly its restricted size and connectivity, and the specific tasks they are good at, has prompted the research on coupling classical and quantum solvers for leveraging the capabilities of both in what is known as hybrid quantum-classical algorithms.  Particularly, \QUBO problems play a special role in \QC, aiming at the minimization of a quadratic objective function over binary variables. Although these problems are NP-hard, they have a simple but quadratic structure and are the standard format for some quantum optimization approaches, such as \QA~\cite{Kadowaki1998}, \QAOA~\cite{Farhi2014} and \VQE~\cite{Peruzzo2014}. Many NP-hard problems, in particular also polynomial \IP problems, can be cast into \QUBO, e.g., by the addition of variables and transferring the constraints into the objective function via penalty terms.

In contrast to gate-model \QC~\cite{Deutsch1989}, quantum annealers~\cite{Johnson2011} are single-purpose solvers, specialized at solving combinatorial optimization problems. The global minimum search is done by evolving the quantum system continuously from an initial state into a final one that encodes the \QUBO solution in what is known as adiabatic quantum computation~\cite{Farhi2000}. Although, current gate-model quantum computers are finding promising applications in machine learning, cryptography, and finance, among others, their utility in solving practical-scale problems is still limited~\cite{Dalzell2023}. 
In contrast, quantum annealers have already shown promising applications in many areas, including power systems~\cite{Abbas2023}. Among the companies developing these devices, \mbox{D-Wave} stands out as the first to release a commercially available quantum annealer in 2011, providing access through their software development kit, \textit{Ocean}~\cite{Dwave2024}. Despite their success, quantum annealers face inherent limitations. Their heuristic nature does not guarantee finding the optimal solution, and hardware constraints require problems to be mapped onto the device’s qubits through a computationally demanding process known as embedding, which determines a mapping between problem variables and the hardware qubits. These limitations of current quantum hardware, and quantum annealers in particular, motivate the research presented in this work. More details are provided in~\autoref{sec: QA}.

In this project, an energy system investment planning problem formulated as a \MILP is used as test case. One of the main challenges in the energy sector is responding to and mitigating climate change. Decarbonization pathways are actively studied \cite{Papadis_2020}, and the growing role of renewable energy sources is reshaping system operation. Consumers increasingly act as prosumers, and decentralized, weather-dependent generation, intermittent loads, sector coupling, and additional storage devices are making energy system models more complex \cite{Ridha2020}. Yet grid infrastructure is not evolving at the same pace \cite{Moreno_2017, Worku_2022}, limiting the full use of renewable sources. Expansion-planning models can address these issues, but higher spatial and temporal detail leads to large, discrete optimization problems that often become intractable for classical solvers. This motivates exploring methods to accelerate their solution.

Large \MILP formulations appear frequently regarding operation and investment optimization in energy systems. In general, \MILP problems are non-convex, which makes it classically not possible to find a solution in polynomial time~\cite{Nemhauser1988}. Throughout the 20th century, several solving methods besides the classical \gls{bnb}~\cite{Basu2022} have been developed to tackle \MILPs, such as \BD~\cite{Benders1962}, Dantzig-Wolfe Decomposition~\cite{Dantzig1961} and Lagrange Relaxation~\cite{Fisher1981}. Each of the aforementioned methods has computational differences depending on the problem structure. But a general approach is to decompose the original problem into smaller and less complex iteratively updated \SPs, i.e., solving simpler problems in expense of repetition. Some constraints, henceforth complicating constraints, have both integer and continuous variables in their expression, which prevent a decomposition of the original problem into two separately solvable problems. But \BD can efficiently manage these complicating constraints enabling the decomposition.

In \BD, the \SPs are coordinated by what is known as a \MP. The \MP and \SPs are solved alternately until a stopping criterion is achieved, thereby approximating the global optimum in every iteration by improving the upper and lower bounds for the solution of the original problem. More precisely, \BD uses the solutions of the \SPs to introduce a set of constraints, henceforth cuts, to restrict the feasible region of the \MP. This motivates the use of \BD for those problems that can be decomposed in such a way that the \SP is a \LP problem, i.e., it has fully continuous variables once the integer variables are fixed, and the \MP handles all the integer variables, which constitute a computational bottleneck. Therefore, techniques for speeding up the solving time of the integer component of the \MILP problems are of major interest. This partly discrete nature of \MILPs encourages the use of heuristic techniques, such as \SA~\cite{Kirkpatrick1983} or \QA~\cite{Kadowaki1998}, to address the integer component, while the continuous \LP part can be efficiently handled using a classical solver ~\cite{Karmarkar1984}.

This work evaluates a quantum version of \BD algorithm which uses \QA to solve the integer component after reformulating it into \QUBO, in the following referred to as \BDQA for short, while the name \QBD refers to any \BD algorithm that integrates the use of quantum computers. The test case is the \TNEP, which focuses on the optimal way to expand the power grid. Its decomposable structure enables the separation of integer variables from the rest, making it well-suited for the application of the \BD algorithm.

\subsection{Related work}

Combining classical and quantum computing resources to form what is called `hybrid solvers` is an active field of research. However, there is not yet a clear consent when and where to incorporate quantum technologies into the solution process. That means, which subroutine is the most promising, to gain the most advantage from the yet limited quantum resources. Hence, further research on this topic is necessary.

There are some recently published studies conducted on the use of quantum computers for the solution of \MILPs under a \BD scheme~\cite{Zhao2022, Ellinas2023, Leenders2023, Barrass2025, Naghmouchi2024, Naghmouchi2025, Paterakis2023}. Depending on which and where quantum technology is introduced in the \BD scheme, there are however different types of \QBD. These include different types of quantum computers to solve the \MP or to decide which cut is added to the \MP from those generated in the \SP at each iteration. In \cite{Zhao2022, Ellinas2023}, a \QBD approach is employed to demonstrate that the considered \MILPs can be solved in competitive time compared with fully classical methods. In both studies, at each iteration, a constraint is added to the \MP, thereby reducing its feasible region and producing a new solution. The work in \cite{Zhao2022} was the first to propose converting the \MP of \BD into a \QUBO in order to use \QUBO solvers to accelerate this part of the algorithm. Although convergence was shown, the evaluation was limited to a proof of concept example with integer coefficients close in magnitude, and no extensive performance analysis was carried out. In addition, the study does not specify which quantum annealer was used. In contrast, \cite{Ellinas2023} provides publicly available tutorial notebooks that demonstrate how to solve specific power system \MILPs using a \QBD framework executed on the D-Wave platform. The study also proposes two techniques for improving the generation of cuts. The first technique selects more informative cuts that improve the solution of the \MP. The second technique takes advantage of the multiple solutions returned by the hybrid solver to solve a set of \SPs simultaneously adding multiple cuts to the \MP in each iteration. The authors did not use a quantum annealer directly. Instead, they rely on a hybrid solver provided by D-Wave Ocean SDK. This simplifies the formulation of the problem but introduces an additional level of hybridization, which makes the actual quantum contribution unclear. As a result, it is not possible to conclude that any improvement observed in either study is due to the direct use of quantum annealing.

In \cite{Leenders2023, Barrass2025} a quantum annealer from D-Wave is used to solve the \MP, together with several strategies intended to improve the efficiency of a \QBD approach. The study in \cite{Leenders2023} presents a proof of concept of a \QBD for the design of multi energy systems. The results show that, for the tested instances, classical computers still achieve better performance than the quantum based method. The work in \cite{Barrass2025} introduces a new technique for generating Benders cuts. The method produces a set of candidate solutions for the \MP through a quantum annealer and then solves the associated \SPs, selecting the cut that corresponds to the most violated \SP. Numerical precision issues are addressed through conservative rounding of the cut coefficients, which reduces the number of slack variables, lowers the qubit requirements, and improves the quality of the resulting cuts. Although this approach enhances the overall process, the study does not examine techniques to reduce the time required to embed the logical problem into the hardware graph, which remains a significant bottleneck.

Other \QUBO solvers can also be employed to tackle the \MP within a \QBD scheme. For instance, in~\cite{Naghmouchi2024}, and later improved in~\cite{Naghmouchi2025}, the authors provide an automated procedure to convert the integer part into a \QUBO format and benchmark the optimization of the \MP using \SA against using a gate-based \QC approach with a neutral-atom quantum processor, on a series of small \MILPs. Their results show that the quantum processor outperforms the classical \SA approach.

In the studies described earlier, the \QBD approaches employ the quantum annealer directly to the \MP, which grows at each iteration as new cuts are added, thus, increasing the qubit demand. In \cite{Paterakis2023}, instead, the \MP is solved classically at every iteration, and the \QA is employed only within a cut-selection subroutine to choose an effective subset of these cuts. A central benefit of this design is that the optimization problem sent to the quantum annealer has a fixed and predictable size, since the cut-selection \SP does not expand as the algorithm progresses. The study demonstrates this hybrid strategy, showing how quantum resources can support multi-cut \BD acceleration without incurring in scalability issues associated with quantum evaluation of the \MP.

Perhaps one of the most serious limitations of those studies that employ the quantum annealer is the consideration of the embedding, which is computationally expensive and thus significantly increases the overall runtime of the \BDQA algorithm. All the mentioned studies use the default heuristic from D-Wave to compute the clique without exploring further possibilities such as using fixed pre-computed embeddings~\cite{lobe2024minor}.

\subsection{Contributions}
Building on the limitations of previous studies, this work proposes a more thorough application of \QA in the \BD scheme. The quantum annealer is used directly to solve the \MP and the cut handling and variable precision is improved with respect to the previously cited references, yielding the use of the minimal number of qubits necessary in each iteration. To the best of our knowledge, this is the first study to incorporate acceleration techniques into the embedding stage of a \BDQA algorithm, and likewise the first publication to comprehensively enumerate and analyze all required steps in deep detail rather than treating the annealer as a black-box solver. This approach, not only provides a general framework for \BDQA algorithms, but also explores the capabilities of a quantum annealer for \MILP problem solving, allowing more precise adaptation to annealer constraints, although the proposed algorithm is hardware-agnostic and can be combined with any \QUBO solver.

In summary, the contributions of this paper are:
\begin{itemize}[leftmargin=1.5em]
    \item An automated and general hardware-agnostic \BDQA algorithm implemented in Python, with several specific improvements on the \BDQA scheme to fully exploit the capabilities of the quantum device and enhance its performance.
    \item The use of precomputed embeddings in the \BDQA algorithm minimizes preprocessing time by avoiding the need to compute the embedding in each iteration.
    \item Scalability, performance, convergence, and time complexity analysis using a quantum annealer for solving the \MP of the \BDQA algorithm under different parameter configurations and embedding strategies.
    \item A generation algorithm to create small and scalable set of \MILP instances of a \TNEP problem with all the required data for investment and operational planning.
\end{itemize}

This work lies at the intersection of classical optimization and quantum computing, applied to energy systems planning. The theoretical elements essential to the study are outlined as follows. \autoref{sec: Theoretical Foundation} introduces the theoretical foundations. \autoref{sec: Methodology} outlines the methodology, including a detailed explanation of how to transform the integer part into a \QUBO, the improvements made to the \BD, the pseudo-code, and the benchmarking procedure. \autoref{sec: Numerical Results} presents the results and lastly, \autoref{sec: Outlook} contains the discussion and the outlook.

\section{Theoretical foundation} \label{sec: Theoretical Foundation}
In this section, the necessary background to understand the methodology is laid out. \autoref{sec: Benders_dec} starts describing how Benders' decomposition algorithm can partition a \MILP problem in two, one of which has only integer variables, thus being easier to translate into \QUBO and having more potential to offer a quantum advantage. \autoref{sec: MILP_to_QUBO} explains how \MILP problems can be translated into \QUBO and \autoref{sec: QA}, explains how the logical graph representing the \QUBO is embedded into the D-Wave hardware-native graph. Finally, \autoref{sec: Transmission Network Expansion Planning} presents the formulation of the \TNEP use case employed in this study.

\subsection{Benders' decomposition} \label{sec: Benders_dec}

In the following, \OP always refers to an arbitrary but fixed \MILP, which can be written as
\begin{equation}
\begin{aligned}
    \min \quad  & \vectorInt^\intercal x + \vectorCont^\intercal y 
    \\
    \textrm{s.t.} \quad & \matrixIntIneq x + \matrixContIneq y \leq \vectorContIneq, 
    \\
                       & \matrixIntEq x + \matrixContEq y = \vectorContEq,
                       \\
                       & x \in \Integer^{n}, ~ y \in \Real^{m}, 
\end{aligned} \tag{OP}\label{prob: MILP}
\end{equation}
where $\vectorInt \in \Real^{n}$ and $\vectorCont \in \Real^{m}$ are the vectors representing the coefficients of the integer variables $x\in\Integer^n$ and continuous variables $y\in\Real^m$ in the objective function~$z(x, y)  = \vectorInt^\intercal x + \vectorCont^\intercal y$. The coefficient matrices $\matrixIntIneq \in \Real^{p\times n}$ and $ \matrixContIneq \in\Real^{p\times m}$ and vector $\vectorContIneq \in \Real^{p}$ form the inequality constraints and the matrices $\matrixIntEq \in \Real^{r\times n}$ and $\matrixContEq\in\Real^{r\times m}$ and vector $\vectorContEq \in \Real^{r}$ form the equality constraints.

\begin{figure}[ht]
\centering
  \includegraphics[width=.6\linewidth]{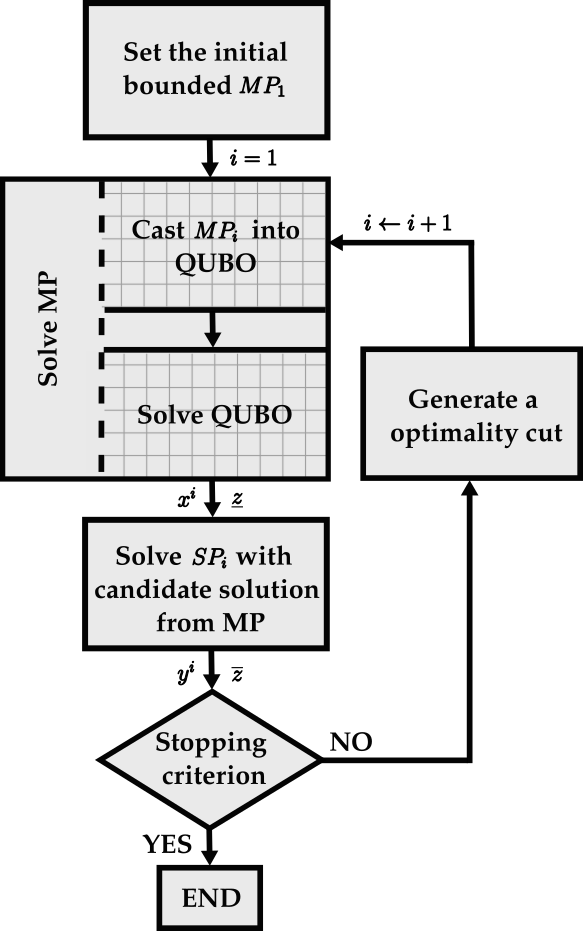}
  \caption{Diagram of the Benders’ Decomposition algorithm. The squared area corresponds to the quantum subroutine of the algorithm.}
\label{fig: benders_diagram_quantum}
\end{figure}

The \BD algorithm~\cite{Benders1962} is a method for solving an optimization problem, by dividing the problem into two smaller problems, the \MP and \SPs, which are updated and solved iteratively, as shown in~\autoref{fig: benders_diagram_quantum}. In this article, the \BD is used to put all the integer variables in the \MP, which is the most challenging part, and produce a linear \SP with continuous variables. Next, the \MP and \SPs are described following~\cite[Sec.\ 6.2]{Conejo2010}.

The subproblem is a simplification of the \eqref{prob: MILP} obtained by just focusing on the inner continuous problem. At iteration $i$, where the integer variables $x$ are fixed to a specific intermediate solution $x^{i}$, it takes the form
\begin{equation}
\begin{aligned}
    \upperVarAux^i = \min \quad  &  \vectorCont^\intercal y\\
    \textrm{s.t.} \quad  & x = x^{i} &&:\lambda,\\
                           & \matrixIntIneq x + \matrixContIneq y \leq \vectorContIneq, \\
                           &\matrixIntEq x + \matrixContEq y = \vectorContEq, \\
                           & y \in \Real^{m},
\end{aligned} \tag{SP$\vphantom{P}_i$} \label{prob: sub}
\end{equation}
where $x^{i}$ are the fixed values for the integer variables and $\lambda$ is the dual variable corresponding to that constraint. If the \SP is unbounded, so is the \eqref{prob: MILP}, and the algorithm terminates. If it is feasible and bounded, the solution to \eqref{prob: sub}, denoted by $y^i$ with the corresponding objective contribution $\upperVarAux^i = \vectorCont^\intercal y^i$, provides together with $x^{i}$ a feasible solution for the \eqref{prob: MILP} and thus an upper bound $\upperBound_i =\vectorInt^\intercal x^{i} + \upperVarAux^{i}$ to the original objective value. Let $\upperBound$ denote the iteratively updated upper bound of the objective function. Note that the integer variables are not fixed arbitrarily but instead are provided by the candidate solution of the \MP. This also means that, if the \SP is infeasible, the \MP is solved again after adding a specific type of constraint that promotes the generation of new solutions $x^{i}$, thus different \SPs in the next iteration. Similarly, a set of constraints are added to the \MP whenever the \SP is feasible, with the aim of promoting better solutions that help to improve the bounds. 

The master problem associated to \eqref{prob: MILP} handles the outer integer contribution and, at iteration $i$, has the form
\begin{equation}
\begin{aligned}
    \min \quad & \vectorInt^\intercal x + \varAux\\
    \textrm{s.t.} \quad 
                        &\varAux \geq  C_j(x)\quad \forall j=1,...,i-1, \\
  &x\in \Binary^{n},~\varAux \in \Real, 
\end{aligned} \tag{MP$\vphantom{P}_i$}\label{prob: master}
\end{equation}
where
\begin{equation} \label{eq:cut}
    C_j(x) = C_{x^j, y^j, \lambda^j}(x) = \vectorCont^\intercal y^j + \lambda^{j\intercal}(x - x^{j})
\end{equation}
is the right hand side of the cuts, added to the \MP, based on the \SP output. The vectors $y^j$ are the collected solutions of the \SPs of all iterations until $i-1$, the vectors $x^j$ are the collected solutions of the \MP, $\varAux$ is a variable that represents a lower bound on contribution of the continuous variables to the objective and $\lambda^j$ is the solution vector of the dual variable to the subproblem constraint $x=x^{j}$ at iteration~$j$, also known as sensitivity. The expression \autoref{eq:cut} follows the approach in~\cite{Conejo2010}, which does not conform to the standard form, but is equivalent and simplifies the cut notation, see \autoref{App: cuts}.

The \MP problem is improved in every iteration $i$ thanks to the subsequent addition of the constraints, known as Benders’ cuts, with function $C_j(x)$ on the right-hand side. If the subproblem is feasible, the added cut is
\begin{equation}
    \varAux  \geq C_j(x),
\end{equation}
which is known as optimality cut, whereas if it is infeasible the added cut is
\begin{equation}
    0 \geq C_j(x)
\end{equation}
which is known as feasibility cut.
The extension by these inequalities in every iteration refines the set of solutions of~\eqref{prob: master}, progressively approximating the original objective value with the correspondingly provided lower bounds $\lowerBound_i =\vectorInt^\intercal x^{i} + \varAux^{i}$. Let $\lowerBound$ denote the iteratively updated lower bound of the objective function.
Note that, although the \MP handles all integer variables of the~\eqref{prob: MILP}, it is not a fully \ILP problem due to the contribution of~$\varAux$.

Furthermore note that, in the first iteration, no cuts are added yet. Since, in this case, $\varAux$ would be unbounded, thereby making the \MP itself unbounded. Therefore, the first candidate integer solution can be calculated by simply optimizing $c^\intercal x$ over $x\in\Binary^n$ as (\MP\!\!$_1$), that means not taking into account $\varAux$ variable in the first iteration. Another option is to manually set a suitable lower bound for $\alpha$ based on knowledge of the problem being optimized.

After solving the \MP and \SP, the following stopping criterion for the global bounds is checked
\begin{equation}
    \upperBound - \lowerBound \leq \varepsilon, \label{eq: stopping}
\end{equation}
for a given desired tolerance $\varepsilon \in \Real_{\geq 0}$.
If it is fulfilled, the algorithm stops and outputs the best-found solution, i.e., the one that triggered the stopping criterion. 
Otherwise, a new iteration of the algorithm starts with a new Benders’ cut generated from the \SP's dual variables, which should improve the candidate solution of the \MP, increase the value of $\lowerBound$ and generate an \SP solution that yields a lower value of $\upperBound$. In the subsequent iterations, the gap is going to be reduced until the stopping criterion is achieved. The convergence behaviour of the \BD in this \MILP formulation is guaranteed for deterministic solvers~\cite[Theorem 2.4]{Geoffrion1972}.

However, finding the optimal solution for a problem with a high number of integer variables in a reasonable amount of time is challenging for classical approaches. In the worst case, every integer solution of the subproblem might need to be explored when such a search algorithm is employed, which leads to a worst-case complexity of the size of the search space, i.e., exponential complexity with the number of integer variables.

\subsection{From (M)ILP to QUBO}\label{sec: MILP_to_QUBO}

In order to use a quantum annealing to solve an \ILP problem, it needs to be transformed into a \QUBO which has the following standard form
\begin{equation}
\begin{aligned}
    \min \quad &x^\intercal Q x\\
    \textrm{s.t.} \quad &x\in \Binary^{n}. 
\end{aligned}
\tag{QUBO}\label{prob: QUBO}
\end{equation}

Note that this is equivalent to the Ising problem, although in the physics community an affine transformation mapping $0$/$1$-variables to $+1$/$-1$ ones is used~\cite{lucas2014ising}.

To obtain an unconstrained problem from a general \ILP of the form
\begin{equation}
\begin{aligned}
    \min \quad  & \vectorInt^\intercal x \\
    \textrm{s.t.} \quad & \matrixIntIneq x \leq \vectorContIneq, \\
                       & \matrixIntEq x = \vectorContEq, \\
                       & x \in \Integer^{n},
\end{aligned} \tag{ILP}\label{prob: ILP}
\end{equation}
the equality constraints can be incorporated into the objective by adding them with a positive penalty term and corresponding positive weight $P_{1}$.
A suitable penalty term ensures that any violation of the equality constraints increases the objective value. Generally,
\begin{equation}
    P_1 \left(\matrixIntEq x - b\right)^\intercal \left(\matrixIntEq x - b\right)
\end{equation}
can be used, as it provides a non-zero addition to the objective, if the constraint is not fulfilled, while it adds nothing, if it is fulfilled. When minimizing, fulfilling the constraint is thus preferred.

To perform the same operation with the inequality constraints, they must be transformed into equality constraints first. This is done using additional variables~$s$, known as slack variables:
\begin{equation}
\begin{aligned}
    \min \quad  & \vectorInt^\intercal x + P_{1}\left(\matrixIntEq x - b\right)^\intercal \left(\matrixIntEq x - b\right)\\
    \textrm{s.t.} \quad & \matrixIntIneq x = h + s, \\
                       & x \in \Integer^{n},~ s \in \Integer^{p}.
\end{aligned}\label{eq:no-ineq}
\end{equation}
Handling these equations analogously as for the equality constraints yields the unconstrained integer optimization problem
\begin{equation}
\begin{alignedat}{2}
    \min \quad  & \vectorInt^\intercal x 
        &&+ P_{1}\left(\matrixIntEq x - b\right)^\intercal \left(\matrixIntEq x - b\right) \\
        &&&+ P_{2}\left(\matrixIntIneq x - h - s\right)^\intercal \left(\matrixIntIneq x - h - s\right) \\
    \textrm{s.t.} \quad & \mathrlap{x \in \Integer^{n},~ s \in \Integer^{p}.}
\end{alignedat}\label{eq: QUO}
\end{equation}

If the variables $x$ and $s$ are already purely binary, i.e., constrained to take values in the set $\{0, 1\}$, the problem is already a \QUBO and can be expressed in the standard formulation by expanding the quadratic terms, expressing the objective in matrix-vector form and renaming all the variables to $x$. 
Actually, once a problem is cast into a \QUBO any solution is feasible from the \QUBO perspective and only a good penalty term is going to increase the chance of getting closer to an optimal and feasible solution from the original problem perspective. However, choosing proper penalty weights for arbitrary constraints can be a hard problem itself.

It is very important to note that for the transformation from \eqref{prob: ILP} to \eqref{eq:no-ineq}, $s$ can only be an integer vector, if the parameter matrix $G$ and the vector $h$ are previously also restricted to contain only integers. To actually represent the same problem, $s$ needs to encode all possible values in the range of the inequality and therefore needs to have the same precision as the parameters.
In case the inequalities do not only contain integer parameters but reasonable rational values, this could be achieved by scaling the whole constraint. 
However, a discretization with a specific precision for the possible values of the slack variables as well as for the parameters is often needed. 
This is in general the case in \eqref{prob: master}, as the \SP of the \BD might return arbitrary dual values for the Benders’ cuts.
The same applies when dealing with continuous variables in \MILPs.
Of course, the quality of the solution can be influenced heavily by that.  
Note that, when not rounding the parameters equally according to the precision of the variables, 
not all possibilities within the constraint range are handled equally. In particular, an assignment that results in a value that deviates from the precision is penalized with a non-zero contribution. This is again a trade-off, which needs to be considered for each use-case individually.

For non-binary variables, a binary encoding is necessary to reformulate the problem as a \QUBO. 
In general, a variable $0 \leq s \leq{} K \in \Natural$ can be represented through a binary expansion:
\begin{equation}
   s = \sum_{i=0}^{k-1} 2^i t_i + (K - 2^k + 1)t_k = \mathbbm{2}_K^\intercal t,
\end{equation}
where $t \in \Binary^{k+1}$ are binary variables for $k = \lfloor \log_2 K \rfloor$.
By stacking the powers of~2 needed to represent an integer~$K$ in the vector $\mathbbm{2}_K$, allows simplification to the right-hand side.
For an interval with non-zero lower bound, i.e., $\underline{K} \leq s \leq{} \overline{K}$ with $ \underline{K}, \overline{K} \in \Integer$, a shift needs to be introduced , and for a specific  precision $\frac{1}{p}$ with $p \in \Natural$ for the discretization of a non-integer variable, the encoding needs to be scaled, such that it results in
\begin{equation}
   s = \tfrac{1}{p} \mathbbm{2}_{K}^\intercal t + \underline{K} \label{eq: binary}
\end{equation}
with $K = p\left(\overline{K} - \underline{K}\right)$.

Notice that the number of necessary binary variables depends on $K$, in a way that the larger the intervals and the higher the precision the more additional variables are needed. In turn, this indicates that representing unbounded variables, and consequently unbounded inequality constraints, is not possible.

In case $x \in \Binary^n$ in \eqref{prob: ILP} (or the \MILP \eqref{prob: MILP}), a special case of a (M)ILP problem is obtained.
However, a (mixed-)binary version cannot be obtained from an arbitrary (M)ILP unless all integer variables are bounded, i.e., $\underline{x} \leq x \leq \overline{x}$, which then allows us to apply the transformations described above. Note that the resulting size of the binary vector $n$ is usually magnitudes larger than it was for the integer version. By the transformation also the parameter vectors and matrices change. However, for simplicity, the same notation is used.

Furthermore, with regard to the small size of current quantum computers, the encoding step might increase the number of binary variable prohibitively. Hence, it is sometimes advisable to rewrite the original optimization problem conveniently or even relax it. See~\cite{glover2019} for further details on the topic. Usually, it requires a trade-off between the number of newly introduced variables and the desired solution quality.

To use \QA to solve the \MP, the reformulation of the \MP into a \QUBO~needs to be incorporated in the \BD scheme, as illustrated in~\autoref{fig: benders_diagram_quantum}.

\subsection{Embedding into the hardware-native problem}\label{sec: QA}
In D-Wave machines, not every qubit interacts with the rest because of connectivity limitations. The graph representing the problem, referred as logical graph, must be embedded into the hardware graph. Specifically, the problem must be mapped into the broken hardware graph since some qubits or couplings of the hardware graph are taken offline because they do not behave as expected after calibration. A vertex in the logical graph, also called a logical vertex, is mapped to multiple qubits, referred to as physical vertices, in the hardware graph, with the requirement that the resulting subgraph making up each physical vertex is connected. For every edge in the logical graph, there must be at least one edge linking the corresponding subgraphs of the two associated logical vertices. The problem of finding such a graph embedding for a problem is NP-hard~\cite{lobe2024minor} and thus computationally expensive. For this reason, heuristics are generally employed to compute a reasonable minor embedding of the logical graph~\cite{Choi2011}. Those are however not guaranteed to succeed and, nevertheless, constitute a computational bottleneck.

Alternatively, a user can precompute the largest complete graph embedding for the hardware graph and use that as a template for mapping the actual problem graph into it, as proposed in~\cite{lobe2021embedding}. Although this is still a hard problem to solve, it shifts the computational effort away from the user. The major drawback of this approach is that the number of logical qubits might be reduced further because of limited connectivity and the imposition of using the complete graph. Moreover, a user has to precompute the largest complete graph embedding frequently due to change in the available qubits and couplings after calibration.

All physical qubits representing a single logical vertex, together forming a logical qubit, are connected by strongly coupling them~\cite{choi2008minor}. Each binary variable of the optimization problem is represented by a logical qubit, that generally is represented in the embedding by more than one physical qubits, due to hardware connectivity. This has multiple drawbacks. First, the number of logical variables that can be mapped onto the machine is drastically limited due to the embedding overhead. Second, the parameters of the embedded problem need to be chosen with great care to really represent the original problem~\cite{lobe2023optimal,lobe2022combinatorialthesis}.
Physical qubits that form a logical qubits should act as one, meaning that all are either 0 or 1 depending on the state of that logical qubit after measuring to be able to de-embed the solution. However, this is not guaranteed and in many situations there are differences between the physical qubits values that represent the logical qubit, this is known as a chain break. In D-Wave, fewer chain breaks correspond to higher confidence in the obtained solution. 

Finally, the coefficients of the \QUBO must respect the resolution limits of D-Wave for both linear and quadratic terms. Due to these precision limitations, large problems requiring high precision may fall outside the practical scope of the quantum solver. Moreover, the errors in the coefficients are not linear, and the scaling of coefficients further affects solution accuracy, see~\cite{dwave_errors} for further details.

\subsection{Transmission network expansion planning}
\label{sec: Transmission Network Expansion Planning}

The \TNEP problem is a \MBLP with binary variables determining which transmission lines are to be build to meet future demand under technical constraints and minimizing the total costs which accounts for investment and operational costs. The formulation provided includes, investment, operational and load shedding costs in the objective function. This problem is extensively used on the base of different future scenarios to achieve carbon neutrality. To facilitate understanding of the problem and the notation used throughout this work, the main variables and symbols are summarized in~\autoref{table:TEPNomenclature}. The units for the variables are MW and the cost is in \euro$/$MW.

\begin{table}[H]
\footnotesize

\begin{tabular}{p{0.05\linewidth} p{0.8\linewidth}}
\multicolumn{2}{c}{\textbf{Greenfield TNEP Nomenclature}}\\
    \hline 
    $\mathcal{N}$ & Set of buses of the network\\
    $\mathcal{C}$ & Set of candidate transmission lines\\
    \hline 
    $x_{kl}$ & Binary decision variable of building a transmission line between buses $k$ and $l$ for $kl \in \mathcal{C}$\\
    $\vectorShedding_{k}$ & Continuous variable for load shedding at bus $k \in \mathcal{N}$ \\
    $\vectorGeneration_{k}$ & Continuous variable for generation at bus $k \in \mathcal{N}$\\
    $\vectorFlow_{kl}$ & Continuous variable for power flow in candidate line $kl \in \mathcal{C}$ \\
    \hline 
    $B_{n, kl}$ & Incidence matrix from line $kl \in \mathcal{C}$ at bus $n\in \mathcal{N}$\\
    $\vectorGenerationUpper_{k}$ & Maximum generation at bus $k \in \mathcal{N}$ \\
    $\vectorDemand_{k}$ & Demand at bus $k \in \mathcal{N}$ \\
    $o_{k}$ & Marginal cost of generation at bus $k \in \mathcal{N}$ \\
    $l_{k}$ & Cost of shedding load at bus $k \in \mathcal{N}$\\
    $\vectorFlowUpper_{kl}$ & Maximum power flow in candidate line $kl \in \mathcal{C}$ \\
    $\vectorInt_{kl}$ & Cost of extending the capacity of the transmission line $kl \in \mathcal{C}$\\
\hline\\
\end{tabular}
\caption{TNEP problem notation.}
\label{table:TEPNomenclature}
\end{table}

This work tackles a greenfield \TNEP, starting from a discrete network with certain topology available and deciding which lines to build in a scenario with a high renewable energy share. Each bus can host gas and/or renewable generators, and the objective minimizes both investment and operational costs. Since renewable sources are cheaper to operate, their presence influences line selection. Investment and operational costs contribute differently to the objective depending on snapshot count. In particular, when considering long-term scenarios, the operational cost dominates, since the cost of transmission lines is spread over many years. In contrast, short-term scenarios tend to favor building cheaper transmission lines at the expense of using more expensive generators. An annualization factor is introduced to balance this for a single snapshot.

The \TNEP is NP-hard~\cite{Oertel2013}, and its complexity increases with the network size. Although there are not short-time constraints, its structure is similar to time-sensitive problems like optimal transmission switching~\cite{Fisher2008}, where fast methods are preferred making the results of the paper potentially transferable.
The \TNEP problem is modeled following~\cite{Dilwali_2016}, which includes the investment costs and a load-shedding term that accounts for unmet demand at a given bus. In this work, the operational cost term is included in the objective function,
\begin{equation}
    z(x,y) = \vectorInt^\intercal x + \vectorCont^\intercal y,
\end{equation}
where $\vectorInt^\intercal x$ represents the investment cost of building lines and $\vectorCont^\intercal y=  o^\intercal \vectorGeneration + l^\intercal \vectorShedding$ with $y^\intercal = (\vectorGeneration^\intercal, \vectorShedding^\intercal)$ and $\vectorCont^\intercal = (o^\intercal, l^\intercal)$, represents the sum of operational costs of producing energy with a generator and the load shedding costs of not fulfilling the demand. The problem has the following set of constraints:
\begin{itemize}[nosep, leftmargin=1.5em]
    \item the power balance constraints
    \begin{equation}
        B\vectorFlow + \vectorGeneration + \vectorShedding = \vectorDemand,
    \end{equation}
    where the incidence matrix $B$ describes how the buses are connected by the lines,
\begin{equation}
B_{n,kl} =
\begin{cases}
+1, & \text{if } kl \in \mathcal{C} \text{ originates from } n\in \mathcal{N}, \\
-1, & \text{if } kl \in \mathcal{C} \text{ terminates at } n\in \mathcal{N}, \\
0,  & \text{if } kl \in \mathcal{C} \text{ not connected to } n \in \mathcal{N},
\end{cases}
\end{equation}
    \item the line capacities
    \begin{equation}
       \lvert \vectorFlow\rvert \leq \vectorFlowUpper x,
    \end{equation}
    \item the generation capacities
    \begin{equation}
       \vectorGeneration \leq \vectorGenerationUpper,
    \end{equation}
    \item the load shedding bounds
    \begin{equation}
        \vectorShedding \leq \vectorDemand,                          
    \end{equation}
    \item with the continuous variables positively defined
    \begin{equation}
        \vectorGeneration {}\in \Real^{\mathcal{N}}_{\geq 0},\, \vectorShedding {}\in \Real^{\mathcal{N}}_{\geq 0},
    \end{equation}
    \item and binary variables for the expansion decision
    \begin{equation}
        x\in \Binary^{\mathcal{C}}.
    \end{equation}
\end{itemize}
Note that, in this implementation of the problem, all possible binary variable assignments are feasible, i.e. there exists an assignment for each continuous variables such that the full solution is feasible, due to the load shedding variable. Therefore, all subproblem in the \BD have feasible solutions and 
thus the cuts generated by the \BD are always optimality cuts.

\section{Methodology} \label{sec: Methodology}
Our hybrid quantum-classical Benders’ Decomposition algorithm, \BDQA, is described in~\autoref{sec: bdqa}, the adaptation of the \TNEP problem to fit in the algorithm is described in~\autoref{sec: tnep_adaptation}, the details of the hardware and software are in~\autoref{sec: Software_Hardware}, the embedding strategies to enhance the algorithm are examined in~\autoref{sec: embedding}, the solver configurations are described in~\autoref{sec: solving_config} and the metrics for benchmarking are described in~\autoref{sec: Benchmarking}.

\subsection{Adjustments in the hybrid quantum-classical Benders’ decomposition scheme}\label{sec: bdqa}

Previously, the general \BD algorithm and the steps to transform an \MILP to \QUBO were described and illustrated, see~\autoref{fig: benders_diagram_quantum}. This section brings these concepts together to develop the \BDQA algorithm.

The \MP of our Benders’ variant is reformulated as a \QUBO, which includes the discretization of the variable~$\varAux$ as described before, which in turn requires the choice of an upper bound $\overline{\varAux}$ and a lower bound $\underline{\varAux}$. Those could be calculated from a worst case estimation of the cut bound(s) or chosen arbitrarily in advance, because they are unset in the first iteration of the scheme. 
Additionally, each cut also needs to include its own slack variables to be transformed: 
\begin{equation}
\begin{aligned} \label{eq:slack}
    &\varAux - C_j(x) \geq 0 \\
    \Leftrightarrow \quad& \varAux - C_j(x) = s^j
\end{aligned}
\end{equation}
for suitable $\underline{s}^j \leq s^j \leq \overline{s}^j$, which also need to be estimated from the cuts or chosen properly. 
An explanation of how these algorithmic choices were addressed in the \BD scheme is provided in what follows.

The \MP \QUBO at iteration $i$ can be written as
\begin{equation}
\begin{alignedat}{2}
    \min \quad  & \vectorInt^\intercal x + \tfrac{1}{p}\mathbbm{2}_{K}^\intercal \beta + \underline{\varAux} \\
        &+\sum_{j = 1}^{i-1} P_{j}\left(\tfrac{1}{p}\mathbbm{2}_K^\intercal \beta + \underline{\varAux}  - C_{j}(x) - \tfrac{1}{p}\mathbbm{2}_{T_j}^\intercal t^j + \underline{s}^j\right)^2 \\
    \textrm{s.t.} \quad & \mathrlap{x \in \Binary^{n},~ \beta \in \Binary^{k+1},} \\
        & \mathrlap{t^j \in \Binary^{\ell_j + 1} ~\forall j = 1, ..., i-1,}
\end{alignedat}\label{eq: MP QUBO}
\end{equation}
with $\beta$ being the binary representation vector of $\varAux$ according to~\eqref{eq: binary} with $K = p(\overline{\varAux} - \underline{\varAux})$ for precision value $p \in \Natural$, analogously with $t^j$ being the binary representation of the slack of constraint $s_j$ with $T_j = p(\overline{s}^j - \underline{s}^j)$ and $\ell_j = \lfloor \log_2 T_j \rfloor$, and suitable choices of a penalty weight $P_j$, for each cut respectively iteration~$j$.

\bigskip


In this work, a variation on the cut added to the \MP is done to improve the performance by reducing the number of required slack variables. Recall that, in every iteration of the algorithm, a new cut is generated. The generated cuts from the \SP \eqref{eq:cut} can be rewritten as
\begin{equation}
\varAux  - \lambda^{j\intercal}x \geq \vectorCont^\intercal y^j - \lambda^{j\intercal}x^{j} = \underline{s}^j,
\end{equation}
where $\underline{s}^j$ is the lower bound of the cut constraint at iteration $j$. Because inequality constraints need to be converted into an equality constraints for the \QUBO formulation of the \MP, using a floating-point bound would require too many additional slack variables due to the required precision. Therefore, this work adopts integer precision for the binary encoding and use $p=1$, meanwhile keeping the parameters $\lambda$ unrounded, with the trade-off of not exactly encoding the constraint values within the bounds, as mentioned before. However, the bounds are conservatively rounded to an integer in each iteration, i.e., to 
\begin{equation}
    \eta^j = \left\lfloor \underline{s}^j \right\rfloor,
\end{equation}
such that valid solutions of the \MP are not excluded. 

In order to provide tight bounds on the required slack variables (cf. \eqref{eq:slack}), which are dependent on the bounds on $\varAux$, the cut constraints are evaluated in the extreme scenarios in our scheme. The worst case is where the decision variables activate either all positive or all negative coefficients of $\lambda^{j\intercal}x$ considering all previous iterations~$j$ with
\begin{equation}
\begin{aligned}
    \underline{\varAux} &= \min \big\{ \big\lfloor\underline{s}^j  +  \lambda^{j}_{-}\big\rfloor \,|\, j = 1, \ldots, i-1 \big\}, \\
    \overline{\varAux} &= \max \big\{ \big\lceil\underline{s}^j  +  \lambda^{j}_{+}\big\rceil \,|\, j = 1, \ldots, i-1 \big\},
\end{aligned}
\end{equation}
where $(\cdot)_-$ and $(\cdot)_+$ take the sums of the negative and positive coefficients, respectively. 
Equivalently, this yields the upper bounds for the inequalities~$\overline{s}^j$.
Note that those are adjusted for all $j = 1, ..., i-1$ in every iteration $i$, like it is done for the bounds of $\alpha$ are.
This enhancement efficiently reduces the number of binary variables required for the binary representation of $\varAux$ and all $s^j$. 
Note that a relaxed upper bound on $\alpha$ was used in favor of obtaining feasible solutions, because when bounds get tight, the heuristic might only choose not only non-optimal but also non-feasible solutions.

However, if a candidate solution that should be excluded by the newly added cut lies too close to the cut, rounding errors may prevent its exclusion, leading to additional iterations.
Candidate solutions from the \MP might be revisited, which implies that a redundant cut with its respective penalty weight and a distinct set of slack variables is added. Notice that this effectively doubles the penalty term associated with the already added cut. This action increases the effective penalty of the constraints and ensures that further iterations are more likely to explore alternative solutions rather than revisiting previously examined ones. Another possible approach would be to not add the cut and hope that the heuristics would find a different solution of the \MP or, alternatively, increasing the penalty of the existing cut without adding it again generating an additional set of slack variables. This, however, is not explored in this work.

In the proposed formulation, single cuts are added at each iteration using a common penalty coefficient. The penalty weight is provided by the user and is fixed and identical for all cuts, as the relative importance of individual cuts is not known \emph{a priori}. For the computational study, the penalty parameter $P$ was selected empirically to obtain a setting that performs robustly across most instances, balancing the improvement of the objective value with the enforcement of constraint satisfaction.

In our \TNEP instances, the dominant contribution to the total cost is the operational cost, which corresponds to the objective of the \SP and therefore to the variable $\varAux$ in the \MP \QUBO{} objective. The same scaling behaviour applies to the constraint term,
\begin{equation}
    \sum_{j = 1}^{i-1} 
    \left(
        \tfrac{1}{p}\mathbbm{2}_K^\intercal \beta 
        + \underline{\varAux}
        - C_{j}(x) 
        - \tfrac{1}{p}\mathbbm{2}_{T_j}^\intercal t^j 
        + \underline{s}^j
    \right)^2
    \;\propto\; \varAux^2 .
\end{equation}
Since the objective term is proportional to $\varAux$ while the constraint term is 
proportional to $\varAux^2$, a reasonable choice for the penalty is therefore $P\propto 1/\varAux$.

To determine a robust and problem-independent penalty weight, $\varAux$ was first replaced by the optimal value obtained classically using \GRB on a representative instance. This value served as an initial guess for the magnitude of the penalty coefficient. All test instances considered in this study share the same total demand, although the demand is distributed differently across nodes, which makes this initial value a meaningful baseline for setting $P$.

Subsequent manual tuning indicated that a better penalty choice was possible, although this refined setting is not included in the computational results presented here. A comprehensive analysis of penalty-weight tuning strategies for \MILP to \QUBO reformulations lies beyond the scope of this work, but the proposed worst-case-based scaling offers a simple and robust heuristic.

As a summary, a scheme of the \QBD algorithm can be seen in~\autoref{alg:QuantumSingleCutBenders}. Notice that, in the algorithm, only optimality cuts are added since the instances considered in this study produces feasible subproblems as stated in~\autoref{sec: Transmission Network Expansion Planning}.

\begin{algorithm}[h]
\SetKwComment{Comment}{/* }{ */}
\SetKwInOut{Input}{input}
\SetKwInOut{Output}{output}
\SetKwProg{Fn}{function}{:}{end}
\caption{\mbox{\texttt{Quantum Benders’ Decomposition}}}
\label{alg:QuantumSingleCutBenders}

\Input{Data of the \MBLP ~(cf. \eqref{prob: MILP}), $q_{\max}$, $\varepsilon \geq 0$}
\Output{Optimal $(x^\ast,y^\ast,\upperBound,\lowerBound)$}

\BlankLine
$(\text{MP}_1)$, $\underline{\varAux}^1$, $\overline{\varAux}^1 \gets $ Initialize from data\;
$i \gets 0$\;

\While{$(\upperBound-\lowerBound) \geq \varepsilon$}{
    $i \gets i + 1$\;
    \If{$i > 1$}{
        {\color{gray}/* Add the Benders’ cut constraint to the previous master problem */} \\
        $(\text{MP}_i) \gets (\text{MP}_{i-1}) \cup (\varAux - (\lambda^{i-1})^\intercal x \geq \mathrlap{\eta^{i-1})
        ;}$
    }
    
    \BlankLine
    $(\text{BLP}_i) \gets $ Discretize $(\text{MP}_i)$ with $\underline{\varAux}^i, \overline{\varAux}^i$\;
    $(\text{QUBO}_i) \gets $ Convert $(\text{BLP}_i)$ to \QUBO\;
    
     \If{\upshape \#variables in $(\text{QUBO}_i)$ > $q_{\max}$}{
         \textbf{break}\;
     }
    
    $(x^i,\underline{\varAux}^i, \lowerBound^i) \gets $ \texttt{Solve}($(\text{QUBO}_i)$)\;
    $(\text{SP}_i) \gets $ Fix $x^i$ in \SP\;
    $(\lambda^i, y^i, \upperVarAux^i, \upperBound^i) \gets $ Solve $(\text{SP}_i)$\;
    $\eta^i \gets \left\lfloor \vectorCont^\intercal y^i - \lambda^{i\intercal}x^{i} \right\rfloor$
    $(\lowerBound, \upperBound) \gets (\lowerBound^i, \upperBound^i)$\;
}
\Return{$(x^i,y^i,\upperBound,\lowerBound)$}\;
\BlankLine

\Fn{\upshape\texttt{Solve}($(\text{QUBO})$)}{
    $(\text{EP}) \gets $ Embed problem $(\text{QUBO})$ in \rlap{hardware\;} \\
    $S \gets $ Sample $(\text{EP})$ with QA\;
    $S' \gets $ De-embed $S$\;
    $(x^*,\underline{\varAux}^*, \lowerBound^*) \gets $ Extract from best sample of $S'$\;
    \Return $(x^*,\underline{\varAux}^*, \lowerBound^*)$\;
}
\end{algorithm}

Notice that the penalty term should be at the same order of magnitude as the objective so that there is not a dominant term but a balance between optimizing and fulfilling the constraints. Based on that, a common penalty value was fixed, which works well for the majority of the cuts. Furthermore, the lower bound~$\lowerBound$ is obtained by omitting the additional term in the objective of the \MP \QUBO formulation, namely the constraint penalization. This is achieved by classically evaluating the solution obtained for the variables
$x$ and $\alpha$ in the \MP objective.

Finally, a stopping criterion terminates the procedure if the gap is less than or equal to $5\%$ or if the maximum allowable \QUBO size $q_{\max}$ is reached, which was set to $160$. This is close to the maximum number of vertices in a complete graph that can be embedded in the current hardware.

\subsection{TNEP reformulation}\label{sec: tnep_adaptation}

A set of small enough test cases solvable on limited quantum hardware
is created through \texttt{PyPSA} Python package~\cite{PyPSA2018}. The openly accessible 37 buses network, \textit{elec\_s\_37}~\cite{horsch_2020}, with all required data for investment planning is selected from \texttt{PyPSA-EUR} dataset~\cite{PyPSAEUR_DATA2023}. The network is reduced by clustering the grid into a given number of buses using a hierarchical agglomerative clustering method included in the aforementioned \texttt{PyPSA} Python package. The generators are reduced to two types, gas and renewable, where renewable sources have cheaper costs. This implies that in some scenarios an expensive line might be build in order to use a set of renewable sources. Originally, the resulting clustered network from \texttt{PyPSA} offers a new topology with certain existing lines connecting the buses. Those lines are removed from the network to achieve what is known as a green model formulation. Network generation provides small enough use cases with all the required data and different topologies to optimize. 

In order to surpass the low bit resolution of \mbox{D-Wave}, see~\cite{dorband2018} for more details, the set of \MILP 
instances are scaled by modifying the monetary and energy units, so that the ratio of \QUBO coefficients
are as close to one as possible having a comparable influence between the linear and quadratic terms of the \QUBO. This enhances the representation with the available precision of D-Wave Advantage 5.4. Our experiments indicate that rounding the coefficients has minimal impact on the solution, since the costs of renewable energy sources are substantially lower than those of gas-based sources. Moreover, the investment planning coefficients have a higher magnitude compared to the operational coefficients. However, a thorough analysis of the effects of rounding the coefficients requires further investigation to fully validate the assumption and it is beyond the scope of the present study.

\subsection{Software and hardware}\label{sec: Software_Hardware}

The \BDQA is implemented in Python and made available in the \texttt{quacla} library~\cite{sergio2026}. The package relies mainly on two Python packages. First, the \texttt{quark} library~\cite{windgaetter2025quantum, lobe2025quarksoftware}, which is used to reformulate the \ILP into \QUBO. Second, \texttt{cvxpy}~\cite{diamond2016cvxpy}, which is used as a modelling language to implement the optimization problems, for the reductions of the \MILP into the corresponding \MP and \SP, and as an interface to classical solvers. This is needed fundamentally to solve the linear \SP in each iteration of the \BDQA, but it is also used to solve the original problem and the \MP for benchmarking purposes.

All classical computations were performed on an AMD Ryzen™ 5 5600X processor, while the quantum computations were performed on a D-Wave Advantage System 5.4~\cite{Boothby2020} with more than 5000 physical qubits with a 15-way connectivity, Pegasus architecture~\cite{dattani2019}. The access to the D-Wave hardware was provided by JUPSI~\cite{JUPSI}.

\subsection{Embeddings}\label{sec: embedding}

Each time an optimization problem has to be solved by the quantum annealer, it must be mapped into the hardware architecture of D-Wave Advantage 5.4. Consequently, if the \BDQA has $r$ iterations, the embedding process must be executed $r$ times. This process involves finding complete subgraphs, also known as cliques, and because computations involving cliques are highly resource-intensive~\cite{lobe2024minor}, precomputing them can substantially reduce the preprocessing overhead. Two embedding strategies are considered in this paper to reduce computational time while maintaining solution quality. Both are compared against the default embedding strategy by D-Wave:
\begin{enumerate}[leftmargin=1.5em]
    \item \texttt{\MM}: It is a heuristic tool for minor embedding and the default embedding strategy of D-Wave. Given a minor and target graph, it tries to find a mapping that embeds the minor into the target~\cite{cai2014, gualmez2025} (D-Wave default).
    \item \texttt{\FX}: It uses the Advantage 5.4 QPU from D-Wave for the \MP and a precomputed 160 complete graph for all \QUBO instances.
    \item \texttt{\TT}: It uses the Advantage 5.4 QPU from D-Wave for the \MP and the tightest precomputed complete graph matching the size of the \QUBO currently being solved.
\end{enumerate}
In this work, a detailed analysis is presented of the numerical techniques employed to reduce qubit requirements, along with a thorough examination of the embedding process, which is a preprocessing step for mapping the logical graph onto the hardware graph. To the best of the authors' knowledge, this is the first work to incorporate pre-computed embeddings to reduce the computational cost of \BDQA optimization.

In quantum annealing on D-Wave hardware, the solver does not return a unique ground-state configuration but instead produces a distribution of energy samples. A \QUBO is solved with \texttt{READS}, generating one candidate solution per read. The solutions are stored in a sample set that records the bit-strings, their number of occurrences, and metadata such as chain breaks. Consequently, practical \QA workflows require substantial classical post-processing: samples must be de-embedded, chain breaks must be resolved, energies recomputed, and the full sample set typically resorted to determine which configuration should ultimately be selected. As emphasized in recent benchmarking work~\cite{tasseff2024}, this post-processing introduces nontrivial overhead that must be considered alongside with the annealing time.

After the samples were obtained from the quantum annealer, they were de-embedded. This includes the handling of ``broken'' samples by employing the default strategy from D-Wave which is majority voting. Afterwards, all returned samples are sorted by their computed energy, and the lowest-energy configuration is selected as the solution to be continued with.

\subsection{Solving configurations}\label{sec: solving_config}

The solver configurations are presented in~\autoref{tab:solver_configurations}, specifying the solver used for the \MP and \SP, the embedding, when applicable, and the corresponding solver parameters.
The optimization with \GRB in the original \MILP formulation is used as a comparison with a state-of-the-art solver. In this case, \GRB has full knowledge of the entire problem and can fully leverage its advanced features and optimizations. The remaining configurations use all \GRB for the \SP, and a specific solver for the \MP. The solvers range from \GRB, to see the actual performance of the \BDQA algorithm compared to state-of-the-art techniques, a simulated annealer provided by D-Wave in~\cite{Dwave2024SA}, and the D-Wave quantum annealer Advantage System 5.4~\cite{Boothby2020}. In the case of using Advantage System 5.4, the whole algorithm is run for the 3 embedding strategies described in~\autoref{sec: embedding}. For each case, the parameters $\texttt{RUNS}$ and $\texttt{READS}$ are fixed. The first parameter corresponds to the number of runs, that is, the number of times the Benders’ algorithm is executed for a given problem. The second parameter specifies the number of reads, i.e., the number of times the quantum annealer samples the \MP in each iteration within a run. The solver parameters are left at their default values unless otherwise specified. In the case of \GRB, for the original \MILP and \MP it uses the Barrier~\cite{Nesterov1994} method and in the \SP it uses the concurrent method, which uses both the simplex algorithm~\cite{dantzig1951} and the Barrier method simultaneously until one outperforms the other, with convergence tolerance of $1\mathrm{e}{-10}$. Moreover, specific heuristics have been developed internally by \GRB to enhance even further the mentioned classical approaches.

A solution is considered successful if its relative error compared to \GRB's optimal solution is less than $5\%$. For each instance, the first 10\% successful solution are selected for benchmarking if available. Otherwise, that instance is considered as failed for the given solver configuration.
Notice that only the objective values are compared, not the individual variable values. It may occur that a solution with feasible variable values leads to unexpected objective values in the \MP due to a wrong value on the optimization of $\varAux$ or the slack variables. To address this, the variable assignments are evaluated classically in the original \MILP, yielding the corresponding objective values.

In the case of the \SA, the number of reads is the same as for the \QA, and $\texttt{SWEEPS}$, which is the analog of the annealing time in a quantum annealer, is set to be equal to $100$, which is found to be a good enough parameter for most of the instances. A sweep is a tour over all variables, trying an update on each one, spin flip. Hence, more sweeps imply more exploration of the solution space.

With these settings, the performance of the solving algorithm, depending on the type of computer, embedding, and number of runs, can be evaluated. Beyond that, the original \MILP problem will be solved with \GRB. Altogether, this makes 15 computations. In~\autoref{appendix}, others configurations are explored.
\begin{table*}[t]
\caption{Solver configurations for QBD. Further configurations are considered in the \autoref{appendix}}.
\centering
\begin{tabular}{@{}lllll@{}}
\toprule
\textbf{Solvers} & \textbf{Embedding} & \textbf{MP}           &  \textbf{SP}           & \textbf{Parameters}\\
\midrule
GRB & None  & - & - & default \\
\midrule
GRB-GRB & None  & GRB & GRB & default \\
\midrule
SA-GRB & None & SA & GRB & \texttt{READS=100};
\texttt{SWEEPS=100} \\
\midrule
QA-GRB-MM & \texttt{minorminer} & QA & GRB & \texttt{READS=100}\\
\midrule
QA-GRB-TT & \texttt{tightest} & QA & GRB & \texttt{READS=100}\\
\midrule
QA-GRB-FX & \texttt{fixed} & QA & GRB & \texttt{READS=100}\\
\bottomrule
\end{tabular}
\label{tab:solver_configurations}
\end{table*}

\subsection{Benchmarking}\label{sec: Benchmarking}

Benchmarking is carried out using a series of indicators, including the objective values, the \QUBO size, the number of iterations of the algorithm, the success rate, the solve time, which for D-Wave corresponds to the sampling time, accumulated across iterations of both the \MP and \SP for the considered instance and the total elapsed time, measured in Python using the \texttt{time} module, which captures the time from problem setup to solution output, including preprocessing, solving, and post-processing times. For exact solvers, the solve time corresponds to the reported solve execution time.

All instances are solved until either the stopping criterion~\eqref{eq: stopping} is met or the maximum \QUBO size is reached for the pre-computed complete graphs, then the number of iterations, solve and total time, the \QUBO size of the last iteration, and the objective value are gathered.

\section{Numerical results} \label{sec: Numerical Results}
In this section, a comparative analysis of the convergence~\autoref{subsec: convergence}, scalability~\autoref{subsec: scalability}, performance~\autoref{subsec: performance}, and complexity~\autoref{subsec: complexity} of various solvers across a set of use cases of increasing problem size is shown. The configuration with the fixed parameters \texttt{(RUNS=100, READS=100)} is presented for clarity, as it provides a reasonable trade-off between computational time and solution quality. The remaining plots obtained by varying the number of reads and runs can be found in the~\autoref{appendix}. The evaluation considers multiple metrics to capture both computational effort and efficiency. The analysis are done with use cases from 3 to 15 buses, since beyond that, the problems are too big for the quantum computer. However, the computational complexity analysis is done up to 38 buses using the conventional solver and simulated annealing. 

\subsection{Algorithm convergence} \label{subsec: convergence}

\begin{figure}[ht]
  \includegraphics[width=\linewidth]{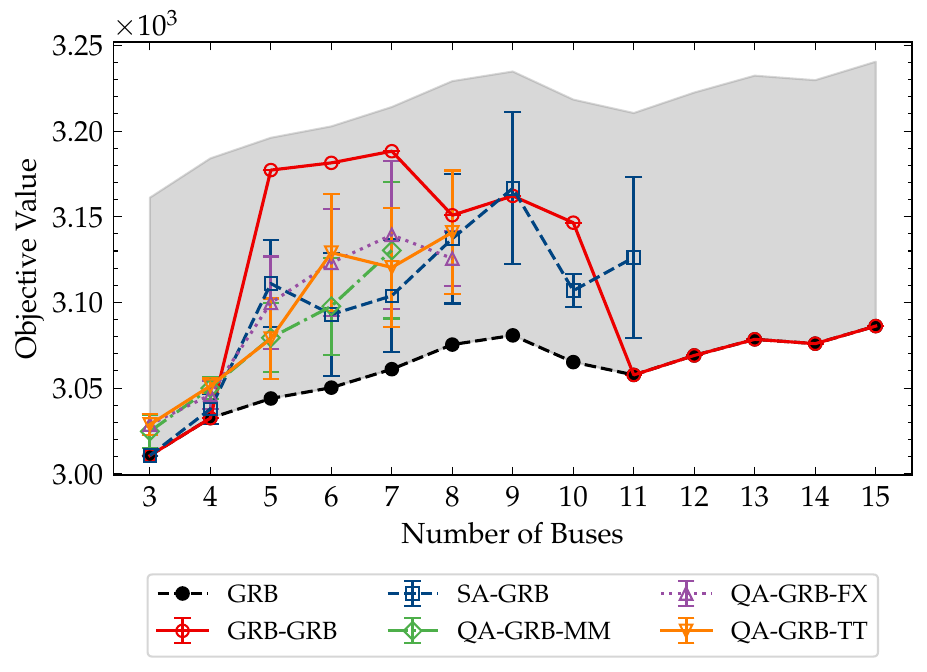}
  \caption{Average objective values compared to the state-of-the-art solution and the 5\% acceptance margin (shaded area). Error bars represent the standard deviation across 10\% successful runs of configuration: \texttt{(RUNS=100, READS=100).}}
\label{fig: convergence_values}
\end{figure}

\begin{figure}[ht]
  \includegraphics[width=\linewidth]{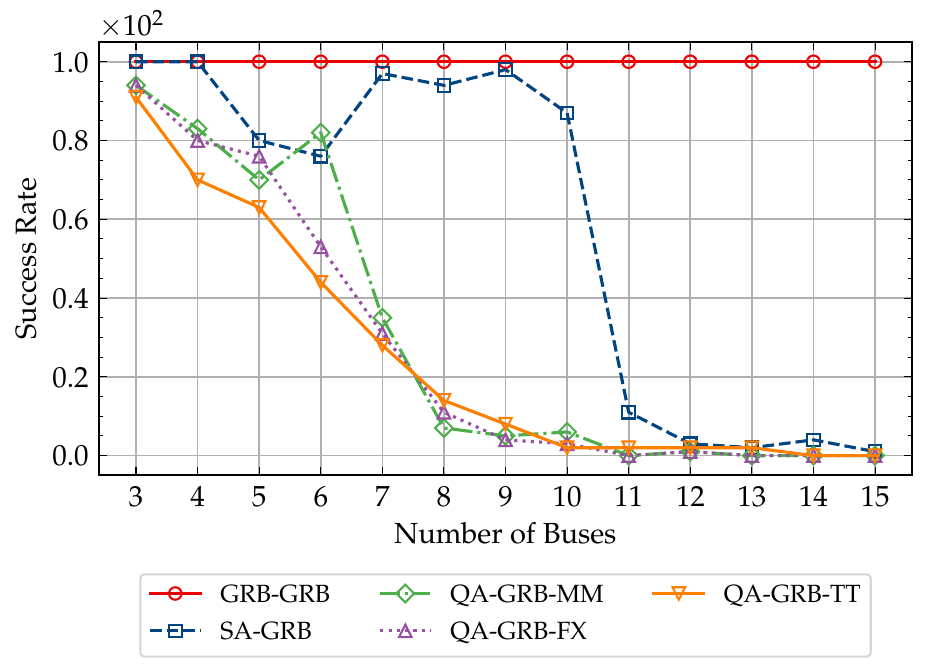}
  \caption{Success rate as function of the number of buses of configuration: \texttt{(RUNS=100, READS=100)}.}
\label{fig: convergence_success}
\end{figure}

The comparison of the objective values against the optimal values produced by the state-of-the-art solver, \GRB, on the original \MILP problem can be seen in~\autoref{fig: convergence_values}. 
Notice that the gray area lies above the optimal value identified by \GRB for the original \MILP, which represents the accepted range for the objective value. Solutions are considered successful only if they lie within $5\%$ of the optimal value obtained by the state-of-the-art solver. Therefore, the objective values obtained by other configurations cannot fall below this bound.
Additionally, the success rate of all runs as a function of problem size is plotted in~\autoref{fig: convergence_success}.
As expected, larger problem instances present greater challenges, resulting in a noticeable decrement in success rates for \SA and \QA while exact solvers maintain a stable success rate, at the cost of needing more time for solving the problem. Since heuristic solvers can disrupt the convergence of the \BD algorithm by producing underestimated upper bounds $\upperBound$ or overestimated lower bounds $\lowerBound$, remedial actions are necessary in order to apply the algorithm to bigger problems, but are not explored in this paper.

\subsection{Scalability}
\label{subsec: scalability}
\begin{figure}[ht]
  \includegraphics[width=\linewidth]{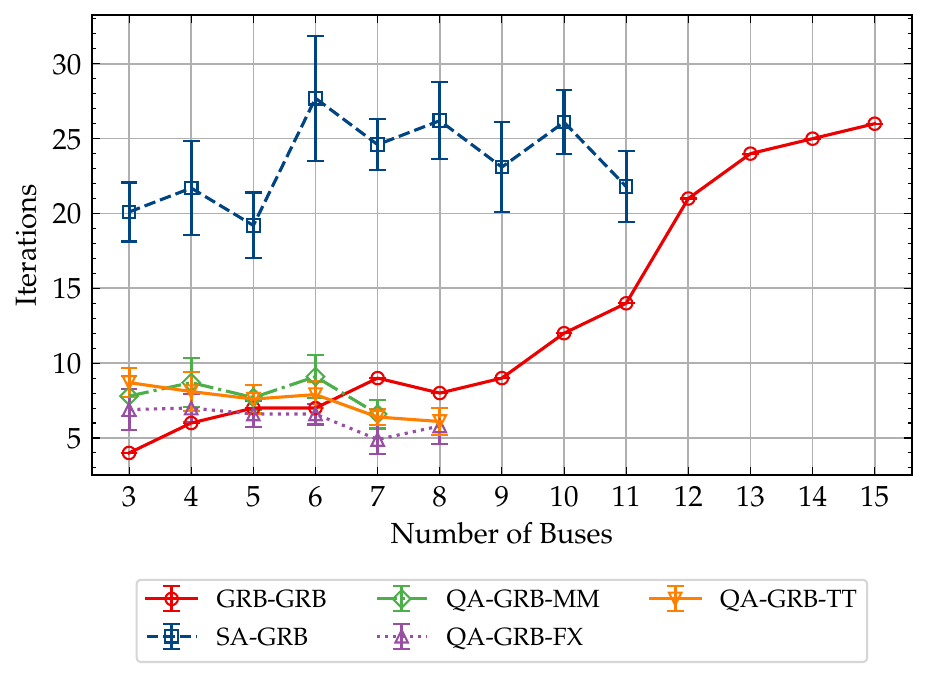}
  \caption{
  Average number of iterations required for convergence as function of the number of buses. Error bars represent the standard deviation across 10\% successful runs of configuration: \texttt{(RUNS=100, READS=100)}.}
\label{fig: scalability_iterations}
\end{figure}

\begin{figure}[ht]
  \includegraphics[width=\linewidth]{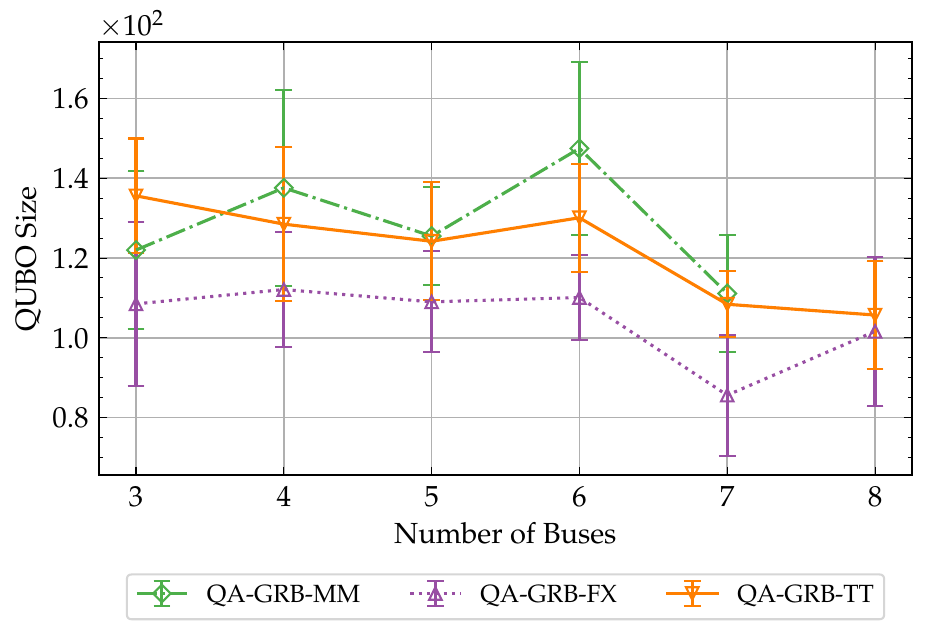}
  \caption{Average (last) QUBO size of each problem instance as function of the number of buses. Error bars represent the standard deviation across 10\% successful runs of configuration: \texttt{(RUNS=100, READS=100).}}
\label{fig: scalability_qubo}
\end{figure}

Each use case solved using \QBD needs a set of iterations for achieving the stopping criterion. While a higher number of iterations would be anticipated for larger use cases, this trend is not evident when employing \SA or \QA from D-Wave for the \MP. In these cases, the iteration count remains largely unchanged, see~\autoref{fig: scalability_iterations}.
In each iteration, a new constraint is added to the \MP, progressively restricting the feasible space. As a result, the size of the \MP increases with every iteration. The largest \QUBO size of each use case is shown in~\autoref{fig: scalability_qubo}.

As expected, the final \QUBO size for \SA is larger than that for \QA with the Advantage 5.4 solver, since it involves a greater number of iterations, which leads to more constraints and ultimately more slack variables. The introduction of additional slack variables increases the \QUBO size at every iteration. Notice that the number of variables of the \QUBO in the last iteration is highly dominated by the slack variables, which explains the relations between~\autoref{fig: scalability_iterations} and~\autoref{fig: scalability_qubo}, more iterations leads to more cuts added to the \MP, which in turn increases the number of slack variables resulting in a larger \QUBO.

\subsection{Performance comparison}
\label{subsec: performance}

\begin{figure}[ht]
  \includegraphics[width=\linewidth]{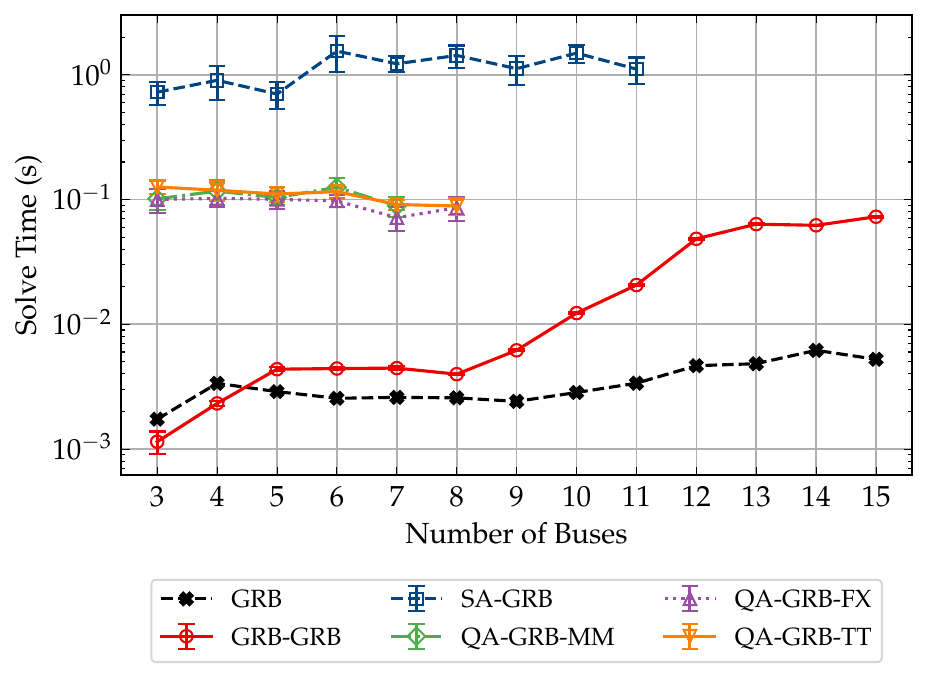}
  \caption{Average solve time (in seconds, in log scale) as function of the number of buses. Error bars represent the standard deviation across 10\% successful runs of configuration: \texttt{(RUNS=100, READS=100)}.}
\label{fig: performance_solvetime}
\end{figure}

\begin{figure}[ht]
  \includegraphics[width=\linewidth]{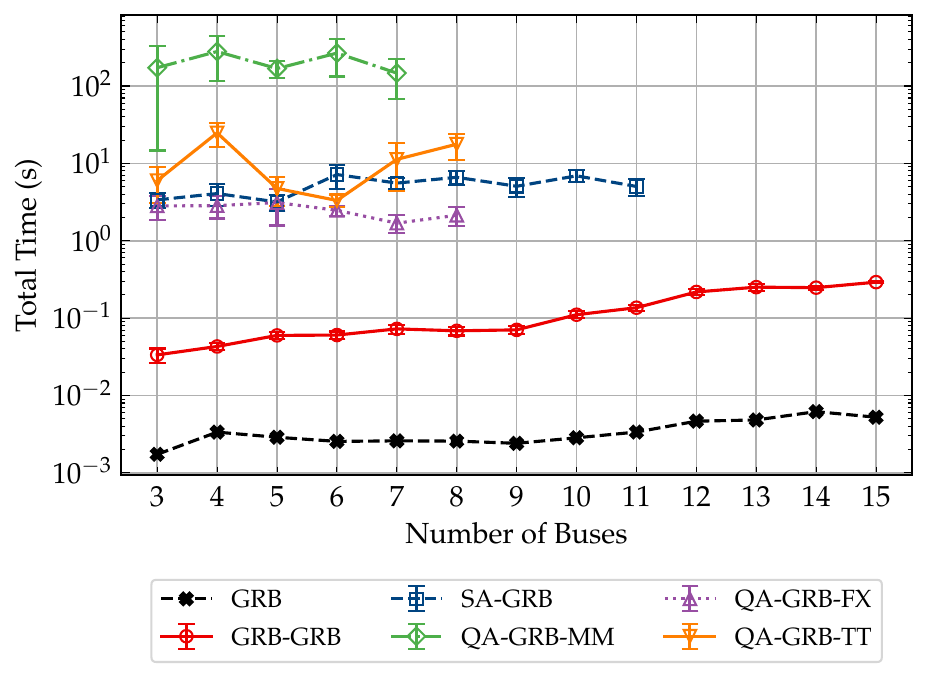}
  \caption{Average total time (in seconds, in log scale) measured the Python \texttt{time} module as function of the number of buses. Error bars represent the standard deviation across 10\% successful runs of configuration: \texttt{(RUNS=100, READS=100)}.}
\label{fig: performance_totaltime}
\end{figure}

In each iteration, a new cut is added, causing the problem to change continuously and requiring parameter re-optimization at every step. Moreover, due to the heuristic nature of \QC, there is no guarantee that the same cuts will be generated across different runs of the use case. Therefore, sufficiently good parameter values that perform well across all iterations were selected, representing a trade-off between solution quality and computational time to achieve a good overall balance.
Solve and total times are plotted as functions of the problem size in~\autoref{fig: performance_solvetime} and~\autoref{fig: performance_totaltime} respectively. Arguably the performance could be explored by optimizing the D-Wave annealer parameters at each iteration, but that would fail to produce a general method.

Notably, for Advantage 5.4, the time required to compute the embedding using the \MM heuristic increases the total time approximately one order of magnitude. As shown in~\autoref{fig: convergence_values}, the \FX and \TT configurations not only achieve similar performance in less time but also succeed in obtaining the correct objective value for larger problem sizes, up to 8 buses compared to 7 buses with \MM. It is not possible to extrapolate from this small sample what the results would look like for larger \QCs, but the key observation is that for \QC using a \FX embedding, the time will depend on the sample time, which can be fixed, and the number of iterations, which is not very different for classical and quantum computers, see~\autoref{fig: scalability_iterations}. Hence, larger problems could potentially be solved faster if the number of reads and annealing time do not increase exponentially. It is also worth noting that \GRB performs better on the original \MILP with its default parameters compared to the implemented \BD. This is not surprising, as \GRB employs various proprietary speed-up techniques to solve the \MILP. When the \MP and \LP are provided separately, these techniques cannot be applied as effectively, since \GRB sees only parts of the problem rather than its full structure.
\subsection{Computational complexity}
\label{subsec: complexity}

\begin{figure}[ht]
  \includegraphics[width=\linewidth]{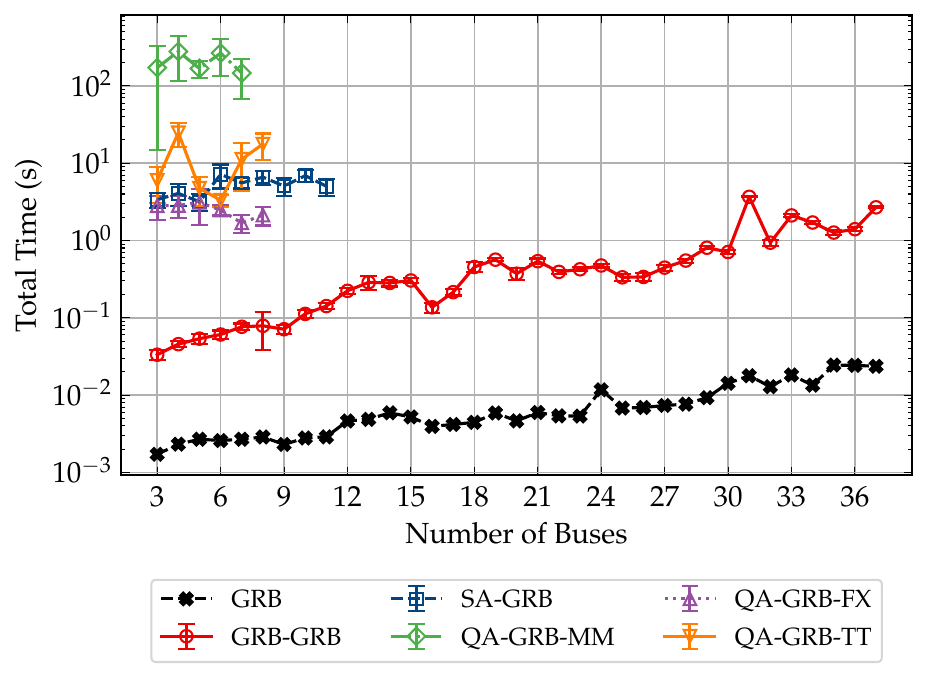}
  \caption{
  Total solution time (in seconds, in log scale) as a function of the number of buses. Error bars represent the standard deviation across $10\%$ successful runs of configuration: \texttt{(RUNS=100, READS=100)}.}
\label{fig: complexity analysis}
\end{figure}

\autoref{fig: complexity analysis} shows the empirical computational complexity based on total observed runtimes across all use cases. This highlights differences between polynomial and potentially exponential trends. For heuristic solvers, runtime scales approximately linearly with the number of iterations and reads, indicating predictable growth. In contrast, exact solvers show non-linear increases, suggesting more complex scaling. \MM embeddings require an order of magnitude more than \FX and \TT embeddings, despite yielding similar solution quality, as mentioned before.

\section{Discussion and outlook} \label{sec: Outlook}
This article benchmarks several configurations, see~\autoref{tab:solver_configurations}, under a \BDQA scheme, evaluating their performance, scalability, and convergence. Previous studies have often overlooked embedding time or have not optimized it using strategies that reduce preprocessing effort at each iteration. In contrast, this work shows that such strategies substantially lower preprocessing time without degrading solution quality.

For the D-Wave Advantage 5.4 quantum annealer, the impact of different embeddings on runtime and solution quality in \MILP optimization under \BDQA is analyzed. In~\autoref{fig: performance_totaltime}, the use of a precomputed embedding strategy, either \FX or \TT, reduces the total runtime of the algorithm by approximately an order of magnitude, particularly in the preprocessing step where the logical graph of the \QUBO must be mapped onto the D-Wave hardware graph. \autoref{fig: convergence_success} confirms that this speedup does not compromise solution quality, even though the embeddings are general rather than problem-specific. Moreover, \autoref{fig: complexity analysis} provides empirical evidence that \BDQA runtime grows linearly with the number of iterations. This is because the \SP is an \LP solvable in polynomial time, and the computational effort of the \QUBO \MP is determined by fixed annealing parameters together with the iteration count. Last, it indicates that quantum solvers remain promising for future applications, although improvements in solution quality are still needed, and larger-scale use cases should be evaluated to strengthen this assumption.

Each iteration of \BDQA adds a single cut to the \MP. Although this restricts the feasible space, the absence of a more effective cut-generation approach and of a cut-selection mechanism leads to redundant cuts, increasing the number of constraints and the slack variables needed to handle them. This limits the size of problems that can be addressed. Additionally, solver parameters remain fixed across all iterations, even though early iterations involve far fewer cuts than later ones. Dynamically adapting these parameters to the evolving \QUBO size could reduce computational effort and improve performance, though this requires further investigation. Also, selecting the lowest-energy sample from the D-Wave output is another simplification, as noted in~\autoref{sec: embedding}, more robust selection rules exist and will be considered. Finally, since the \MP is solved heuristically, convergence of the \BD algorithm is not guaranteed. The heuristic does not guarantee finding the optimal \QUBO solution, and even an optimal \QUBO assignment can be suboptimal for the original \MP due to rounding and precision issues see~\autoref{sec: bdqa}. Thus, remedial actions are necessary in order to apply the algorithm to bigger problems, but are not explored in this paper.

Future research should develop cut-generation strategies that incorporate \QA outputs to produce deeper, more informative cuts, as well as systematic cut-selection methods to reduce redundancy and the associated slack variables. These improvements would increase the scalability of the \BDQA approach on current quantum hardware. Embedding optimization, precision-enhancement methods, and more sophisticated solution-selection techniques also present opportunities to further improve efficiency and solution quality which are key requirements for scaling hybrid quantum-classical optimization to large, real-world \MILP instances.

A complementary research direction is to consider problem classes beyond the convex formulations typically studied. Non-convex and time-constrained optimization problems may offer even stronger opportunities for quantum-assisted heuristics, as classical solvers often struggle to find high-quality solutions within limited time. Exploring these more complex and time-sensitive settings, along with additional enhancement techniques, represents an important avenue for future work.

\section*{Acknowledgements}
The underlying research is part of the projects ‘Algorithms for quantum computer development in hardware-software codesign’ (ALQU), \url{https://qci.dlr.de/en/alqu}, and ‘Applications, interfaces and data formats for quantum computing algorithms in energy system modelling’ (Attraqt'em), \url{https://qci.dlr.de/en/attraqtem},
which were made possible by the DLR Quantum Computing Initiative (QCI) and the German Federal Ministry for Economic Affairs and Climate Action (BMWK). Furthermore, the authors acknowledge the support from the DLR project ‘Enhanced probLEm solVing with quAntum compuTErs’ (ELEVATE).
The authors gratefully acknowledge the J\"ulich Supercomputing Centre (\url{https://www.fzjuelich.de/ias/jsc}) for funding this project by providing computing time on the D-Wave Advantage™ System JUPSI through the J\"ulich UNified Infrastructure for Quantum computing (JUNIQ). The authors also acknowledge the use of the Gurobi Optimizer (Gurobi Optimization, LLC), under an academic license, for solving the optimization problems in this study. 

\section*{Author Contributions}
The core of this work, which includes the main theoretical and code development, analytical calculations, numerical simulations, figure preparation, and manuscript writing, was carried out by Sergio López-Baños. Elisabeth Lobe and Oriol Raventós contributed to the theoretical development, the textual elaboration, the interpretation of results, and the revision of the full manuscript, editing it as necessary. Ontje Lünsdorf contributed to the development of the associated code. Large language model tools were used to assist with language editing. All authors discussed the results and approved the final manuscript, and take full responsibility for the work.

\newpage
\normalsize
\printbibliography

\appendix
\onecolumn
\newpage
\section{Cuts equivalence} \label{App: cuts}

In this section, it is shown that the cuts generated by two equivalent formulations of the linear programming \SP, are identical. First, consider the original \SP written in standard form, \eqref{prob: slp}, by replacing the fixed value of $x$ with $x^{i}$. This removes the need for introducing a constraint to enforce that value.

\begin{equation}
\begin{aligned}
    \upperVarAux^i = \min \quad  & \vectorCont^\intercal y\\
    \textrm{s.t.} \quad &  \matrixContIneq y \leq \vectorContIneq - \matrixIntIneq x^{i} &&:\mu, \\
                           &\matrixContEq y = \vectorContEq - \matrixIntEq x^{i}  &&:\nu, \\
                           & y \in \Real^{m}.
\end{aligned} \tag{SP$\vphantom{P}_i$} \label{prob: slp}
\end{equation}
The corresponding dual problem \eqref{prob: dsp}, is given by,
\begin{equation}
\begin{aligned}
    \upperVarAux^i = \max \quad  &  (\vectorContIneq - \matrixIntIneq  x^{i})^\intercal \mu + (\vectorContEq - \matrixIntEq x^{i})^\intercal \nu\\
    \textrm{s.t.} \quad &  \matrixContIneq^\intercal \mu  + \matrixContEq^\intercal \nu ={}  \vectorCont, \\
                       & \mu \in \Real^{p}_{\leq 0}, \,\, \nu \in \Real^{r},
\end{aligned} \tag{DSP$\vphantom{P}_i$} \label{prob: dsp}
\end{equation}
where $\mu, \nu$ are the dual variables associated with the \SP constraints. 
The resulting solutions are $y^i$, $\mu^i$ and $\nu^i$ and the contribution of the continuous variables to the objective value is thus
\begin{equation}
    \upperVarAux^i = \vectorCont^\intercal y^i
        = (\vectorContIneq - \matrixIntIneq x^i)^\intercal \mu^i + (\vectorContEq - \matrixIntEq x^i)^\intercal \nu^i .
\end{equation}

The corresponding optimality cut is
\begin{equation}
    \alpha \geq (\vectorContIneq - \matrixIntIneq x)^\intercal \mu^i + (\vectorContEq - \matrixIntEq x)^\intercal \nu^i .
\end{equation}
Replacing $x$ by $x + x^i - x^i$ and expanding terms yields to the compact form,
\begin{equation}
\begin{aligned}
    \alpha &\geq (\vectorContIneq - \matrixIntIneq x^{i})^\intercal \mu^i + (\vectorContEq - \matrixIntEq x^{i})^\intercal \nu^i - (\matrixIntIneq(x-x^{i}))^\intercal \mu^i - (\matrixIntEq(x-x^{i}))^\intercal \nu^i\\
    &= \upperVarAux^{i} + (x-x^{i})^\intercal(-\matrixIntIneq^\intercal \mu^i - \matrixIntEq^\intercal \nu^i) \\
    &= \upperVarAux^{i} + (x-x^{i})^\intercal \lambda^i,
\end{aligned}
\end{equation}
with $\lambda^i = -\matrixIntIneq^\intercal \mu^i - \matrixIntEq^\intercal \nu^i$.

In the alternative formulation, $x$ is treated as a decision variable of the \SP, but its value is fixed to the corresponding $x^{i}$ through an explicit constraint,
\begin{equation}
\begin{aligned}
    \upperVarAux^i = \min \quad  & \vectorCont^\intercal y \\
    \textrm{s.t.} \quad  & x = x^{i} &&:\tilde{\lambda},\\
                           & \matrixIntIneq x + \matrixContIneq y \leq \vectorContIneq  &&:\tilde{\mu}, \\
                           &\matrixIntEq x + \matrixContEq y = \vectorContEq  &&:\tilde{\nu}, \\
                           & x \in \Real^{n},\,\,y \in \Real^{m},
\end{aligned} \tag{M-SP$\vphantom{P}_i$} \label{prob: msp}
\end{equation}
where $\tilde{\mu}, \tilde{\nu}, \tilde{\lambda}$ denote the dual variables associated with the constraints of the alternative formulation, \eqref{prob: msp}. Note that in \eqref{prob: msp}, the variable $x$ is relaxed to the real domain to enable the computation of the corresponding dual value.

The corresponding dual \eqref{prob: mdsp} is,
\begin{equation}
\begin{aligned}
    \upperVarAux^i = \max \quad  & \vectorContIneq^\intercal \tilde{\mu} +  \vectorContEq^\intercal \tilde{\nu} + (x^{i})^\intercal \tilde{\lambda}  \\
    \textrm{s.t.}\quad &  \matrixContIneq^\intercal \tilde{\mu}  + \matrixContEq^\intercal \tilde{\nu} ={} \vectorCont, \\
    &  \matrixIntIneq^\intercal \tilde{\mu}  + \matrixIntEq^\intercal \tilde{\nu} + \tilde{\lambda}= 0, \\
                           & \tilde{\mu} \in \Real^{p}_{\leq 0}, \,\, \tilde{\nu} \in \Real^{r}, \,\, \tilde{\lambda} \in \Real^{n},
\end{aligned} \tag{M-DSP$\vphantom{P}_i$}\label{prob: mdsp}
\end{equation}
where the dual objective $\upperVarAux^i$ is the same as the objective of the standard formulation since both problems are equivalent.
Note that the dual variables are denoted with a tilde, indicating that they are not assumed to take the same values as those in the previous formulation.
Let again $\tilde{\mu}^i$, $\tilde{\nu}^i$ and $\tilde{\lambda}^i$ denote the optimal solutions.
Thus, 
\begin{equation}
    \vectorContIneq^\intercal \tilde{\mu}^i+ \vectorContEq^\intercal \tilde{\nu}^i+ (x^i)^\intercal \tilde{\lambda}^i = \vectorCont^\intercal y^i + \vectorInt^\intercal x^i.
\end{equation}

The corresponding cut is
\begin{equation}
    \alpha \geq \vectorContIneq^\intercal \tilde{\mu}^i + \vectorContEq^\intercal \tilde{\nu}^i + x^\intercal \tilde{\lambda}^i.
\end{equation}
Replacing $x = x + x^{i} - x^{i}$ and expanding yields
\begin{equation}
    \begin{aligned}
        \alpha &\geq \vectorContIneq^\intercal \tilde{\mu}^i + \vectorContEq^\intercal \tilde{\nu}^i +  (x^{i})^\intercal \tilde{\lambda}^i + (x-x^{i})^\intercal \tilde{\lambda}^i \\
        &= \upperVarAux^i + (x-x^{i})^\intercal \tilde{\lambda}^i, 
    \end{aligned}
\end{equation}
where, with the second equality constraint of \eqref{prob: mdsp}, the equivalence of the formulations is shown.

\section{Sensitivity analysis} \label{App: sensitivity}
All computations involving the \QC were repeated using different numbers of runs and reads, and the corresponding results are plotted below. Across the datasets, the configuration \texttt{(RUNS=100, READS=100)} offers a balanced trade-off between computational cost and solution quality. Configurations such as \texttt{(RUNS=100, READS=10)} execute faster and still achieve reasonably good objective values, although the lower number of reads slightly reduces the success rate. As quantum hardware continues to improve, it may become possible to use fewer reads without compromising solution quality, further decreasing computational time. Finally, note that error bars are omitted for \texttt{(10,1000)} because 10\% of 10 runs corresponds to just a single measurement, rendering statistical variation meaningless.

\label{appendix}
\begin{figure}[h]
\centering
\begin{tabular}{l|cc}
& \textbf{\quad\quad\, Objective Value vs Problem Size} & \textbf{\quad\quad\, Success vs Problem Size} \\ 
\hline
\rotatebox{90}{\texttt{\quad\quad (RUNS=100, READS=10)}}& \includegraphics[ width=.45\linewidth]{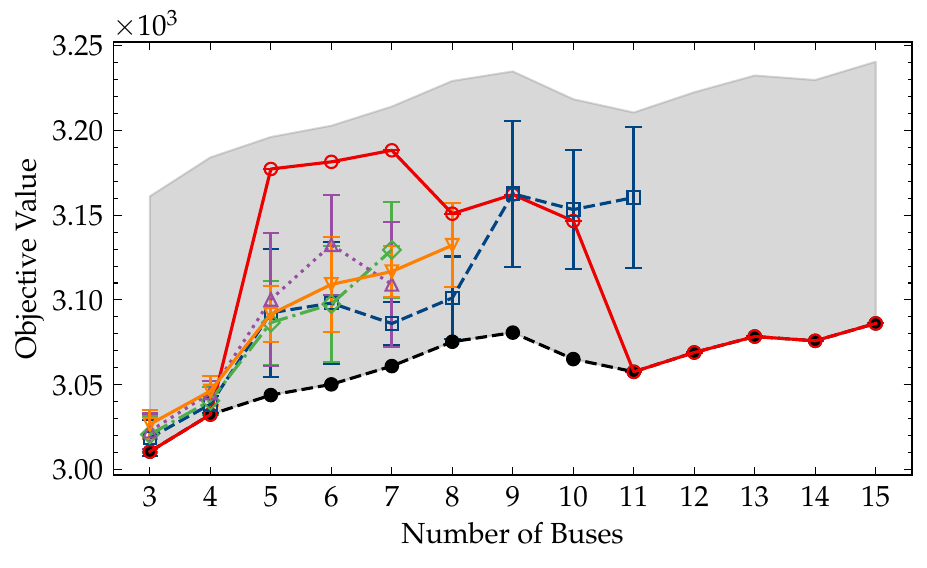} & \includegraphics[ width=.45\linewidth]{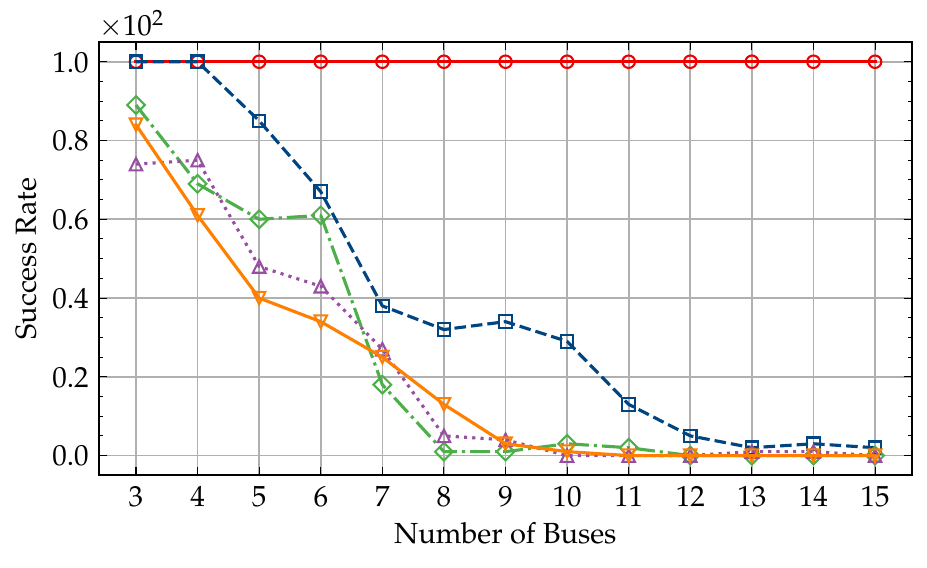} \\
\rotatebox{90}{\texttt{\quad\quad (RUNS=10, READS=1000)}} & \includegraphics[ width=.45\linewidth]{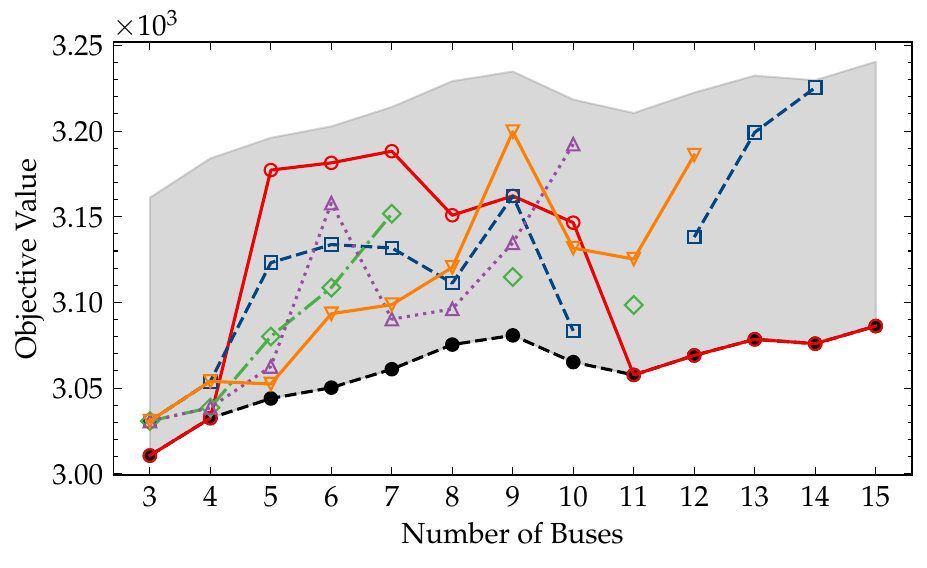} & \includegraphics[ width=.45\linewidth]{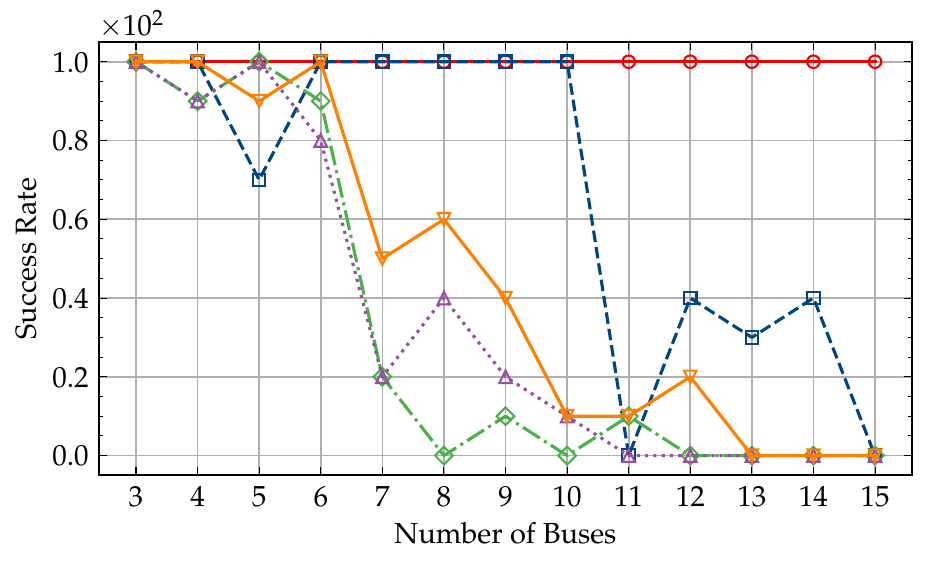}\\
\multicolumn{3}{c}{\hspace{2.5em}\includegraphics[width=0.95\linewidth]{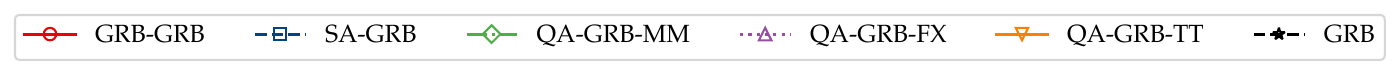}} \\
\end{tabular}
\caption{\textbf{Objective value and success rate as functions of problem size across different solver configurations.}\\
Each row corresponds to a different number of reads and repetitions: (top) 100 runs with 10 reads and (bottom) 10 runs with 1000 reads. The left column shows the average objective values compared to the state-of-the-art solution and the 5\% margin (shaded area), while the right column shows the corresponding success rates. Error bars represent the standard deviation across $10\%$ successful runs.}
\label{fig: app_conv_values_success}
\end{figure}
\newpage
\begin{figure}[h]
\centering
\begin{tabular}{l|cc}
& \textbf{\quad\quad\, Iterations vs Problem Size} & \textbf{\quad\quad\, QUBO size vs Problem Size} \\ 
\hline
\rotatebox{90}{\texttt{\quad\quad (RUNS=100, READS=10)}}& \includegraphics[ width=.45\linewidth]{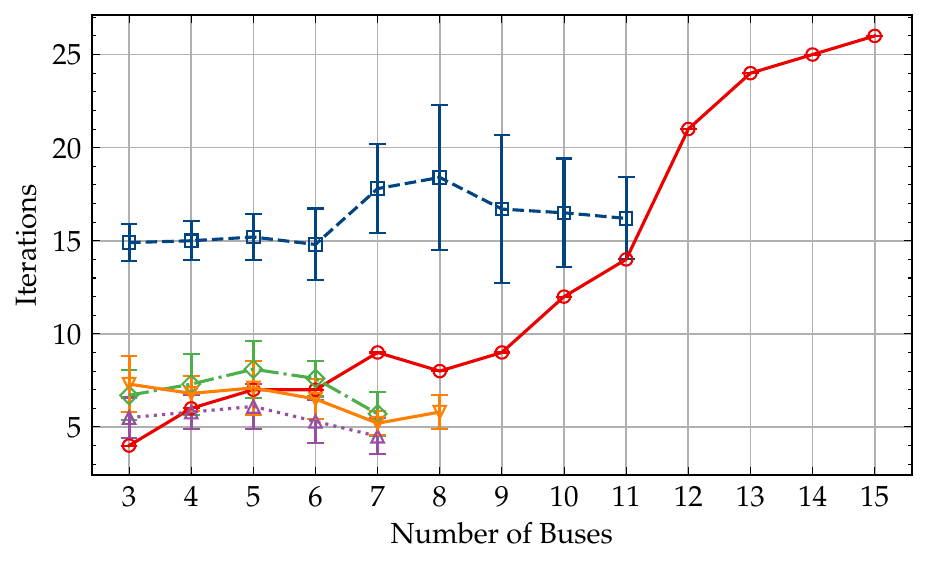} & \includegraphics[ width=.45\linewidth]{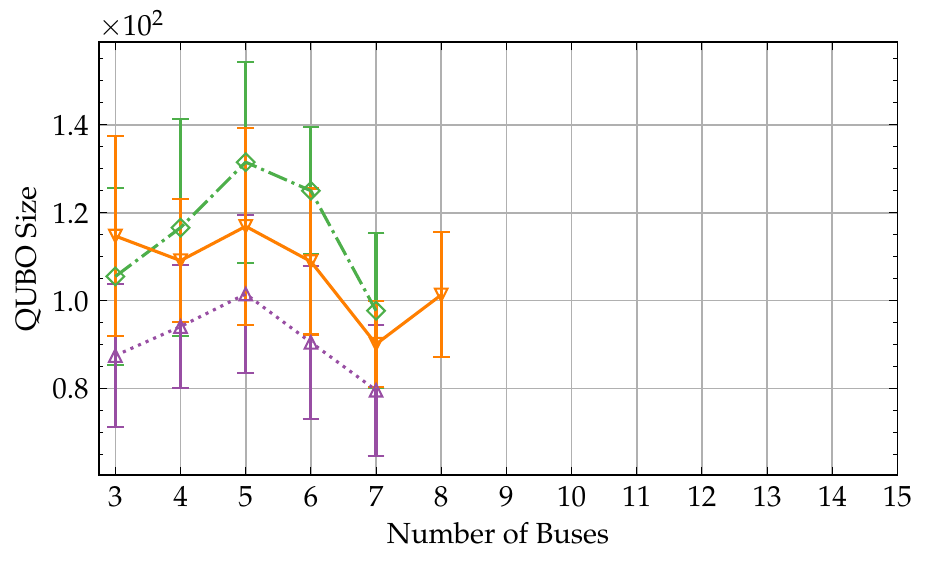} \\
\rotatebox{90}{\texttt{\quad\quad (RUNS=10, READS=1000)}} & \includegraphics[ width=.45\linewidth]{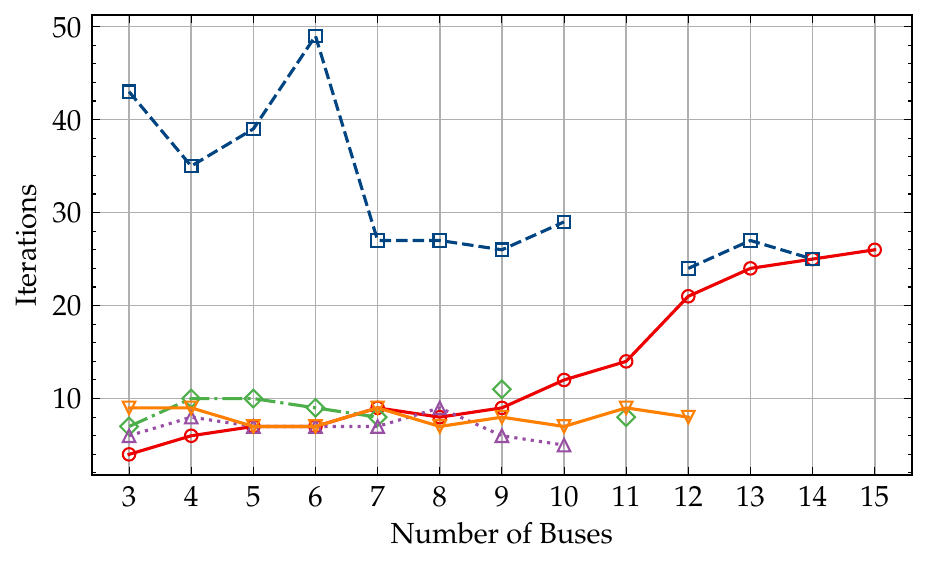} & \includegraphics[ width=.45\linewidth]{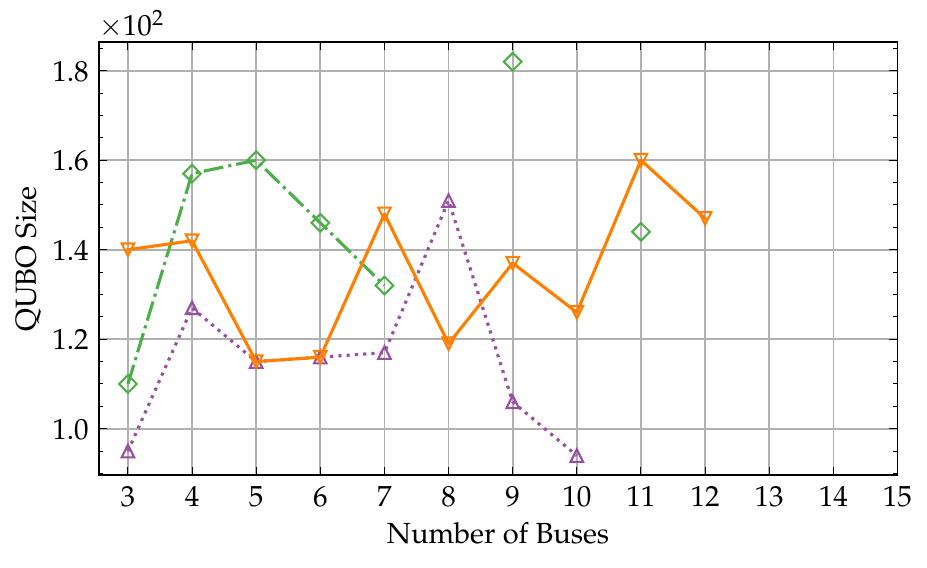}\\
\multicolumn{3}{c}{\hspace{2.5em}\includegraphics[width=0.95\linewidth]{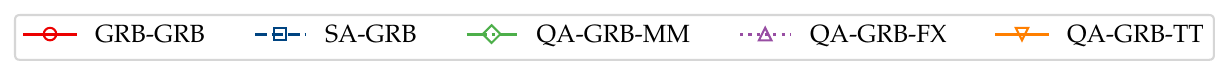}} \\
\end{tabular}
\caption{\textbf{Iteration count and QUBO size as functions of problem size across different solver configurations.}\\
Each row corresponds to a different number of reads and repetitions: (top) 100 runs with 10 reads, (middle) 100 runs with 100 reads, and (bottom) 10 runs with 1000 reads. The left column shows the average number of iterations required for convergence, while the right column reports the last QUBO size of each problem instance. Error bars represent the standard deviation across $10\%$ successful runs.}
\label{fig: app_iter_QUBO_size}
\end{figure}

\newpage
\begin{figure}[h]
\centering
\begin{tabular}{l|cc}
& \textbf{\quad\quad\, Solve Time vs Problem Size} & \textbf{\quad\quad\, Total Time vs Problem Size} \\ 
\hline
\rotatebox{90}{\texttt{\quad\quad (RUNS=100, READS=10)}}& \includegraphics[ width=.45\linewidth]{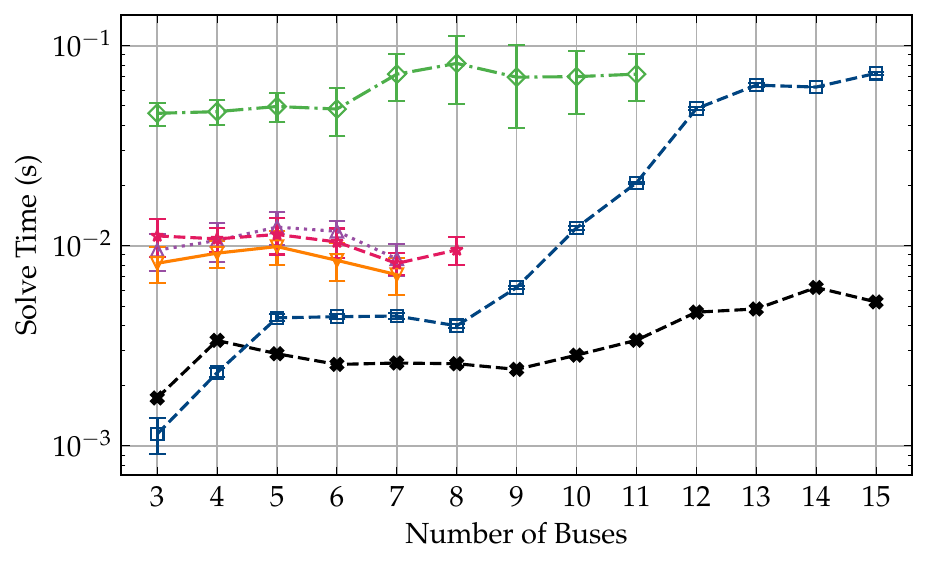} & \includegraphics[ width=.45\linewidth]{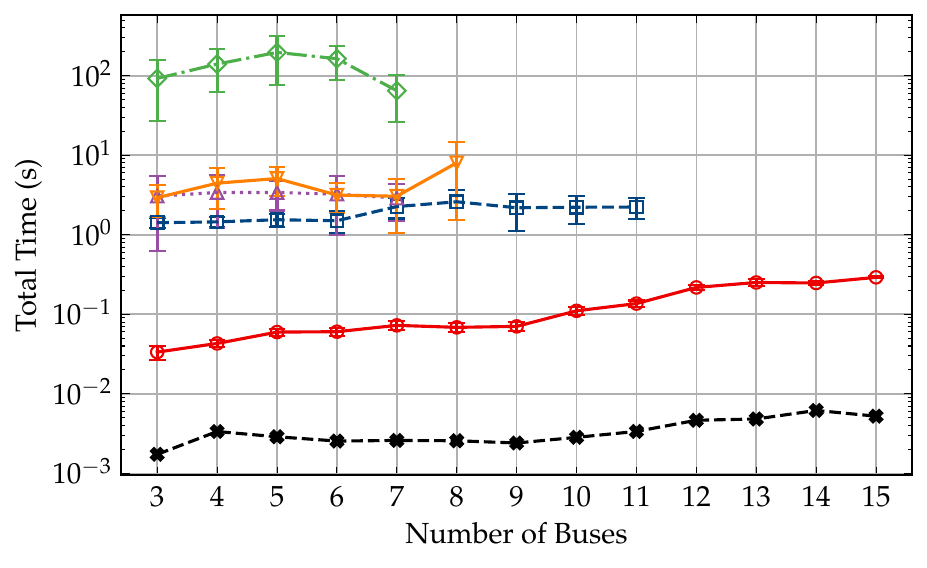} \\
\rotatebox{90}{\texttt{\quad\quad (RUNS=10, READS=1000)}} & \includegraphics[ width=.45\linewidth]{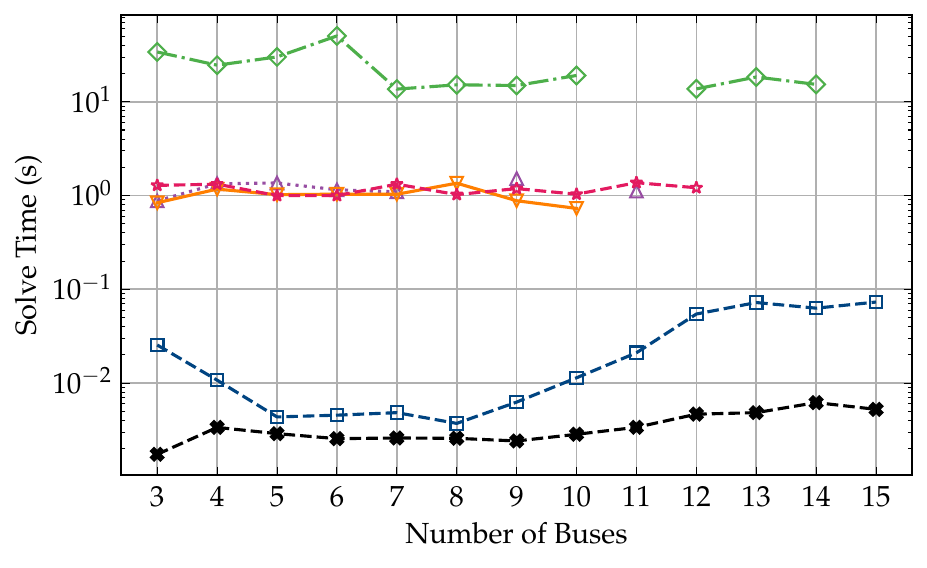} & \includegraphics[ width=.45\linewidth]{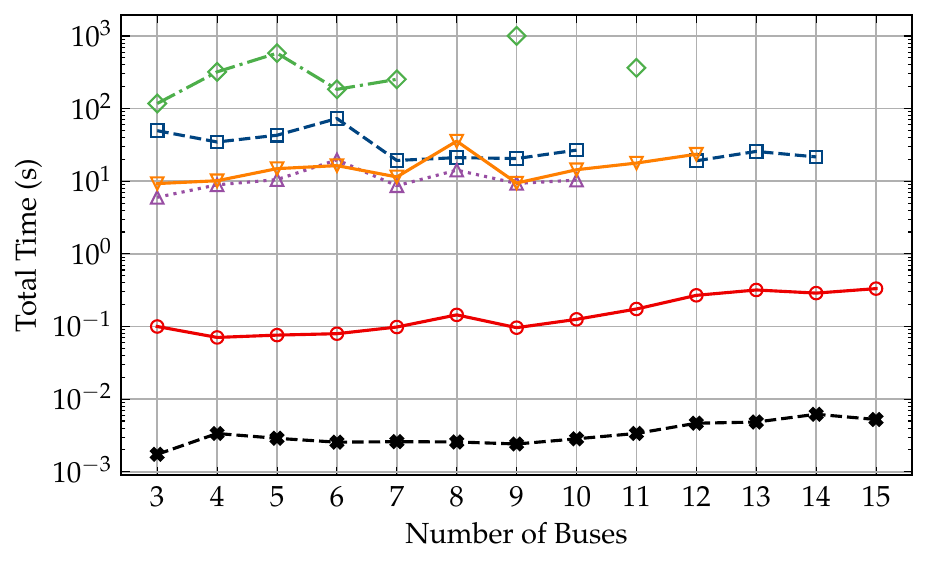}\\
\multicolumn{3}{c}{\hspace{2.5em}\includegraphics[width=0.95\linewidth]{figures/shared_legend_3.pdf}} \\
\end{tabular}
\caption{\textbf{Solve time and total time vs. problem size for various solvers and embedding strategies.}\\
Each row corresponds to a different number of reads and repetitions: (top) 100 runs with 10 reads and (bottom) 10 runs with 1000 reads. The left column shows the solve time (in seconds) as a function of the number of buses, and the right column shows the corresponding total time (in seconds), both plotted on logarithmic scales. Error bars represent the standard deviation across $10\%$ successful runs.}
\label{fig: app_times}
\end{figure}
\newpage

\end{document}